\documentclass[prb,showpacs,amsmath,amssymb,superscriptaddress,twocolumn,floatfix]{revtex4-1}
\usepackage{graphicx, epsfig}
\usepackage{amsmath,amssymb}
\usepackage{graphicx}
\usepackage{color}
\usepackage{epstopdf}
\usepackage{bm}

\newcommand{\diff}{\ensuremath{\mathrm{d}}}

\newcommand{\pwrt}[2]{\ensuremath{\frac{\partial #1}{\partial #2}}}
\newcommand{\spwrt}[2]{\ensuremath{\frac{\partial^2 #1}{\partial #2^2}}}
\newcommand{\bfr}{\ensuremath{\mathbf{r}}}

\newcommand{\bra}[1]{\ensuremath{\textstyle \left\langle #1 
	\right|\displaystyle}}
\newcommand{\ket}[1]{\ensuremath{\textstyle \left| #1 
	\right\rangle\displaystyle}}
\newcommand{\braket}[2]{\ensuremath{\textstyle \left\langle\left. #1 
	\right| #2 \right\rangle\displaystyle}}
\newcommand{\bracket}[3]{\ensuremath{\textstyle \left\langle #1 \right| #2
	\left| #3 \right\rangle\displaystyle}}

\begin{document}

\title{Large Stark Effect for Li Donor Spins in Si}

\author{Luke Pendo}
\author{E.M. Handberg}
\affiliation{Physics Department, South Dakota School of Mines and Technology,Rapid City, SD 57701}
\author{V.N. Smelyanskiy}
\affiliation{NASA Ames Research Center, Mail Stop 269-1, Moffett Field, CA 94043}
\author{A.G. Petukhov}
\email{Andre.Petukhov@sdsmt.edu}
\affiliation{Physics Department, South Dakota School of Mines and Technology,Rapid City, SD 57701}

\date{\today}

\begin{abstract}
We study the effect of a static electric field on lithium donor spins in 
silicon. The anisotropy of the effective mass leads to the anisotropy of the 
quadratic Stark susceptibility, which we determined using the Dalgarno-Lewis 
exact summation method.
The theory is asymptotically 
exact in the field domain below Li-donor ionization threshold, relevant to the 
Stark-tuning electron spin resonance experiments. 
To obtain the generalized Stark susceptibilities at arbitrary fields, we propose a new 
variational wave function which reproduces the exact results in the low-field limit.
With the calculated susceptibilities at hand, we are able to predict 
and analyze several important physical effects. First, we observe that the 
energy level shifts due to the quadratic Stark effect for Li donors in Si  are equivalent to, 
and can be mapped onto, those produced by an external stress. 
 Second, we demonstrate that the Stark effect anisotropy, combined with the unique 
valley-orbit splitting of a Li donor in Si, spin-orbit interaction and 
specially tuned external stress, may lead to a very strong modulation of the 
donor spin $g$-factor by the electric field. Third, we investigate the influence 
of random strains on the $g$-factor shifts and quantify the random strain 
limits and requirements to Si material purity necessary to observe the $g$-factor
Stark shifts experimentally. Finally, we discuss possible implications of 
our results for quantum information processing with Li spin qubits in Si.
\end{abstract}

\maketitle

\section{Introduction}
Recent research has shown that electron spins bound to donors in Si are some 
of the most viable candidates for scalable, solid-state quantum computing 
applications
\cite{Kane:1998wh}.
The lithium donor is unique among other Si shallow donors 
\cite{Smelyanskiy:2005uv} because its ground state is degenerate.~\cite{Watkins:1970wj,Jagannath:1981vc} We will 
demonstrate that the inverted electronic structure of the lithium donor 
allows intriguing possibilities for the efficient manipulation of Li spin qubits using stress and electric fields. The ability
to couple individual electron spins with local electric fields is a key feature of many spin-control proposals for
quantum computing and single-electron spintronic applications. \cite{Kane:1998wh,Loss:1998ia,Vrijen:2000dk,Jiang:2001de,Kato:2003ce,Zutic:2004vi,Petta:2005kn,Tokura:2006iu,Nakaoka:2007jh} 
The electrical
control of single electron spins based on various coupling mechanisms has been demonstrated experimentally for semiconductor quantum dots.~\cite{Nowack:2007du,Laird:2007eq,PioroLadriere:2008hs,NadjPerge:2010kw,Shafiei:2012th}
Also, a possibility of spin-orbit-driven electrical modulation of shallow donor $g$-factors in GaAs has been predicted theoretically.~\cite{De:2009tx}
In this work, we present theoretical evidence that the Li donor in Si is yet another shallow donor 
system that possesses such
functionality. This system may play a significant role in silicon-based quantum information processing (QIP) with spin
qubits.

The Stark effect for shallow donors in silicon has been studied both 
experimentally \cite{Bradbury:2006gz} and theoretically. \cite{Friesen:2005fc, 
Smit:2004fx, Martins:2004ke} Most of the theoretical calculations  have been 
performed for substitutional donors (e.g. P)~\cite{Friesen:2005fc} but not for 
Li donors occupying tetrahedral interstitials in a silicon lattice.~\cite{Watkins:1970wj} 
Furthermore, most of the previous studies of substitutional donors 
ignore the interplay between the Stark and Zeeman effects. This 
interplay occurs because the spin-orbit interaction couples the non-degenerate ground state of a substitutional 
donor to the rest of $1s$ manifold,~\cite{Hasegawa:1960ua,Roth:1960wp} 
which is separated form the ground state  by large 
valley-orbit splitting, $\Delta_{vo}\sim$10 meV. Since the spin-orbit coupling is much smaller than
$\Delta_{vo}$,  the electrical modulation of the  
$g$-factors of substitutional donors is strongly 
suppressed. On the contrary, the ground state of the interstitial Li donor is 
degenerate and even modest spin-orbit interaction may have a profound effect 
on the energy levels leading to a strong electric field dependence of the Zeeman splittings. 

Intriguingly, the orbital degeneracy of the ground state quintet of the 
lithium donor gives rise not only to nontrivial spin-orbit effects
\cite{Watkins:1970wj} but also to a strong long-range {\it elastic-dipole} 
coupling between the orbital states of different lithium donors. 
\cite{Smelyanskiy:2005uv} The elastic-dipole coupling was studied earlier in connection to an acceptor-based 
quantum computing scheme in silicon proposed by Dykman and Golding 
\cite{Golding:2003tq}. Thus, it is of interest to study the lithium donor 
electron as a new candidate for  QIP 
applications with a long-range inter-donor coupling.

Experimental techniques utilizing pulsed electron spin resonance (ESR) 
measurements in interdigitated devices based on Sb-doped Si revealed 
observable Stark shifts of ESR lines in an electric field ${\cal E}$.
The shifts can be parametrized as $\Delta A/A=\eta_a {\cal E}^2$ and $
\Delta g/g=\eta_g {\cal E}^2$, for the hyperfine constant and the $g$-factor 
shifts respectively, where $\eta_a\approx -3.7\times10^{-3}\,\mathrm{\mu 
m^2/V^2}$ and $\eta_g\approx -1\times 10^{-5}\,\mathrm{\mu m^2/V^2}$.
\cite{Bradbury:2006gz} We anticipate that experiments on similar devices 
based on Li-doped Si may reveal substantially larger shifts of the magnetic resonance 
lines on the order of 10 gauss for electric fields ranging from 0 to 3 kV/cm. This proposition 
seems somewhat surprising as there is no hyperfine interaction in the ground 
state for Li donors and the strength of the spin-orbit interaction on the Li 
atom in Si is negligible.~\cite{Watkins:1970wj} Nonetheless, spin-orbit effects have been observed quite 
prominently in ESR spectra of Li-doped natural Si under external 
stress.~\cite{Watkins:1970wj}

\begin{figure}[tbh]
\includegraphics[width=2.5 in]{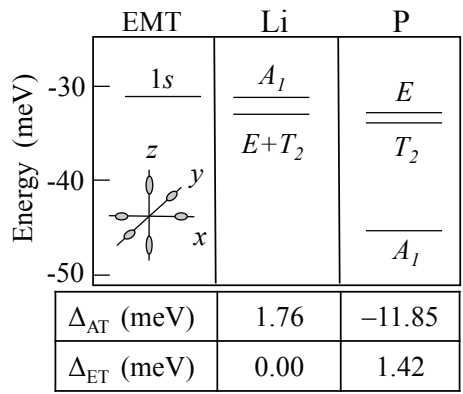}
\caption{Upper table shows the energy level diagrams for $1s$  electron of the 
shallow donors in silicon: left column corresponds to EMT
 calculations giving the binding energy $31.27\,\mathrm{meV}$.~\cite{Faulkner:1969uj} 
Experimental values for the $1s$ energies of 
substitutional phosphorus\cite{Larsen:2003iw} and interstitial lithium 
\cite{Jagannath:1981vc} donors are shown in right  and middle columns, 
respectively. Ionization energy of lithium donor $33.02\,\mathrm{meV}$ is in 
particularly close agreement with EMT result.  Inset depicts the locations of 
6 conduction band minima in silicon. Lower table shows the experimental 
results for the energy differences between the singlet and triplet 
($\Delta_{AT}$) and between the doublet and triplet ($\Delta_{ET}$).
\label{fig:energy-table}}
\end{figure}

This happens because the unique electronic structure of a Li donor enables observation of 
spin-orbit effects which have {\em crystalline} rather than impurity 
origin.~\cite{Watkins:1970wj} The situation in P-doped Si is quite different because
the crystalline spin-orbit interaction 
is much less important due to its weak influence on the isolated, non-degenerate ground state. 
More specifically, the orbital states of the $1s$ manifold of a 
shallow donor in silicon form a singlet $A_1$, a doublet $E$ and a triplet 
$T_2$. For substitutional donors the singlet ground state $A_1$ is separated 
by a gap exceeding $10\,\mathrm{meV}$ from the closely spaced doublet and 
triplet. The sequence of levels is inverted for the interstitial Li donors in 
such a way that the ground state is five-fold  degenerate and composed of the 
orbital doublet $1s(E)$ and triplet $1s(T_2)$ while the singlet $1s(A_1)$ lies 
$1.76\,\mathrm{meV}$ above (see Fig.~\ref{fig:energy-table}).

Within the Effective Mass Theory (EMT) the $1s$ donor electron manifold has 
6-fold orbital degeneracy corresponding to 6 minima of the silicon conduction 
band ${\bf k}_{sj}$. These minima are located along coordinate axes at about 85$\%$ of the 
distance to the zone boundary as shown in the inset of Fig.~\ref{fig:energy-table}:
\begin{equation}
{\bf k}_{sj}\textmd{=}s \kappa_0\,\bm{n}_{j}, \quad s=\pm 1,
\quad  \kappa_0\approx0.85\,{G_{100}}/{2}, 
\label{k0}
\end{equation}
\noindent
where $\bm{n}_{j}$ is a Cartesian unit vector, $j=x,y$ or $z$,  $G_{001}$=
${4\pi}/{a_{0}}$ is a magnitude of the reciprocal lattice vector in [001] 
direction, and $a_0$ is the Si lattice constant. The position of each valley 
is characterized by a composite index $\{sj\}$, where $s=\pm 1$  describes the 
valley centered on either the positive or the negative semi-axis, respectively.  

A short-range tetrahedral potential corresponding to the local symmetry of the 
donor site splits the valley degeneracy.  The intervalley 
effects~\cite{Kohn:1955vn} are described by the valley-orbit Hamiltonian 
\begin{align}
H_{vo}=&\left(E_0+\Delta_{0}\right)\sum_{i,s}|si\rangle\langle si|
+\Delta_{1}\sum_{s,i}|si\rangle\langle -si|\nonumber\\
&+\Delta_{2}\sum_{i,j,s,s^\prime}(1-\delta_{ij})|si\rangle\langle s^\prime j|.
\label{eq:Hvo0}
\end{align}
Here, $E_0$ is the binding energy corresponding to the solution of the
single-valley Coulomb problem and the parameters $\Delta_i$ are the 
matrix elements of  the short-range central cell potential in the basis of the 
six valley orbitals
\begin{equation}
\label{valley-orbital}
\langle{\bf{r}}|sj\rangle=\exp\left(i{\bf k}_{sj}\cdot{\bf r}\right)u_{sj}({\bf r})F_{sj}({\bf r}),
\end{equation}
where $u_{sj}({\bf r})$ is a periodic part of the Bloch function corresponding 
to the center of the valley ${\bf k}_{sj}$, and  $F_{sj}({\bf r})$ is the $1s$ 
envelope function which, along with $E_0$, is found by solving the 
single-valley Coulomb problem. 

The six eigenstates of the Hamiltonian~\eqref{eq:Hvo0}, referred 
to as {\em symmetrized valley orbitals}, can be expressed as 
\begin{equation}
|\mu\rangle=\sum_{s,j}\alpha_{sj}^\mu|sj\rangle, \quad \mu=1,\ldots,6 .
\label{orbeigen}
\end{equation}
Each orbital $|\mu\rangle$ in Eq.~(\ref{orbeigen}) belongs to the irreducible 
representation $\mu$ of the tetrahedral group $T_d$ characterized by the 
valley-orbit coefficients $\alpha_{sj}^{\mu}$ that are given in 
Table~\ref{tab:vo}. 

The singlet state $A_1$ and the doublet states $E_{\theta}$, $E_{\epsilon}$ 
are ``even" with respect to the axis inversion, i.e.  they are symmetric 
combinations of the opposite valley orbitals, $\alpha_{sj}$=$\alpha_{-sj}$. 
On the contrary, each triplet state $T_{2j}$ ($j$=$x,y$ or $z$) is an 
``odd" antisymmetric combination of just two opposite valleys orbitals, 
$\alpha_{sj}$=$-\alpha_{-sj}$.
We  have labeled the symmetrized valley orbitals according 
to Watkins and Ham~\cite{Watkins:1970wj} and indicated their transformational properties under 
the group $T_d$ (last column in Table~\ref{tab:vo}). Thus the singlet, triplet 
and doublet states possess $s$, $p$ and $d$-like characters, respectively.
\begin{table}
\begin{tabular}{|c|c|c|c|}
  \hline
    & $\mu$-\textrm{states}& valley-orbit coefficients \{$\alpha_{sj}^{\mu}\}$& basis functions\\ \hline
 1 & $A_1$ & $\frac{1}{\sqrt{6}}(1,1
,1,1,1,1)$& $x^2+y^2+z^2$ \\[0.5ex] \hline
  2 & $E_\theta$ & $\frac{1}{\sqrt{12}}(-1,-1,-1,-1,2,2)$ &$2z^2-x^2-y^2$\\[0.5ex]  \hline
  3 & $E_\epsilon$ & $\frac{1}{2}(1,1,-1,-1,0,0)$ &$\sqrt{3}(x^2-y^2)$\\[0.5ex]  \hline
  4 & $T_{2x}$ & $\frac{1}{\sqrt{2}}(1,-1,0,0,0,0)$ & $x$\\[0.5ex]  \hline
  5 & $T_{2y}$ & $\frac{1}{\sqrt{2}}(0,0,1,-1,0,0)$ & $y$\\[0.5ex]  \hline
  6 & $T_{2z}$ & $\frac{1}{\sqrt{2}}(0,0,0,0,1,-1)$ & $z$\\
  \hline
\end{tabular}
\caption{\label{tab:vo} Table of valley-orbit coefficients for silicon donor 1$s$ orbital states. 
First column defines the order of states. Second column shows the state label 
derived from the corresponding representation of the group $T_d$. Each row in 
the third column gives a set of 6 coefficients $\{\alpha_{sj}\}$ for $sj$=$\pm 
1,\pm 2, \pm 3$ corresponding to the silicon conduction valley minima at 
$\pm x,\pm y,\pm z$ axes, respectively. Fourth column shows basis functions 
describing the transformational properties of the corresponding states under 
the operations of the group $T_d$}
\end{table}

For phosphorus donors in silicon, Friesen~\cite{Friesen:2005fc} discussed two 
competing mechanisms which determine the behavior of the electron levels in 
the $1s$ manifold. First, as the electric field is increased, the quadratic 
Stark shift causes the energy of each valley to decrease. This will influence 
the form of the valley-orbit Hamiltonian \eqref{eq:Hvo0}. Formally, we have to 
replace $E_0$ with valley-dependent terms $E_{0i} = E_0 - \delta_{i}(
\mathcal{E})$ and place these under the summation sign in Eq.~\eqref{eq:Hvo0}, 
where the corrections $\delta_{i}$ include the quadratic Stark shifts of each 
valley. Secondly, similar substitutions are required for the central cell 
terms in the Hamiltonian $\Delta_0\rightarrow\Delta_{0i}$, $\Delta_1
\rightarrow\Delta_{1i}$ and $\Delta_2\rightarrow\Delta_{2ij}$. These 
parameters $\Delta$ are the matrix elements of the short-range central cell 
potential between the states $|is\rangle$; they are proportional to the 
amplitudes of the envelope functions $F_{sj}(0)$ in the central cell. Since 
the electric field ${\cal E}$ pulls the electron away from the donor site, the 
amplitudes $F_{sj}(0)$ and the matrix elements $\Delta$ are decreasing 
functions of ${\cal E}^2$. As a result, the energy spectrum of the manifold 
narrows and the ground state shifts upward in the direction opposite to the 
quadratic Stark shift.

Our studies reveal that the narrowing effect is not as important for Li as for 
P donors. Because the valley-orbit splitting for Li is considerably smaller 
than phosphorus, the narrowing of the spectrum does not overwhelm the 
quadratic Stark effect in determining the overall behavior of the $1s$ 
manifold in the presence of an electric field. According to our findings, the 
most important effect for Li is the anisotropy of the quadratic Stark effect. 
This anisotropy allows the electric field to induce unique splitting of the Li 
ground state and leads to a very non-trivial interplay of the Zeeman and Stark 
effects.

Our approach utilizes the Dalgarno-Lewis exact summation method to determine 
the quadratic Stark susceptibility. While this calculation is quite lengthy, 
it provides us with an important calibration tool to further 
devise our variational function in the presence of the electric field and 
gauge it against the exact small-field asymptotic behavior. We will discuss how the 
ground state splitting caused by the electric field can replicate 
stress effects and can potentially be used to manipulate a Li spin qubit. 
While most of the theoretical papers on the Stark effect were concerned with 
rather large 
fields,~\cite{Smit:2003hx,Smit:2004fx,Rahman:2007ev,Lansbergen:2008bs} our 
studies are focused on the field domain below the 3-5$\,\mathrm{kV/cm}$ 
relevant to the ESR Stark experiments, i.e. below ionization threshold of 
shallow donors in Si.

The paper is organized as follows. In Section~\ref{sec:Quadratic} we study the 
quadratic Stark effect for a single-valley Schr\"{o}dinger equation for a 
shallow donor in Si. The goal of this section is to calculate the Stark 
susceptibility, which is asymptotically exact in the limit of low electric 
fields using an exact summation method of Dalgarno and Lewis tailored to 
account for the effective mass anisotropy. Based on the findings of 
Section~\ref{sec:Quadratic},  we propose a new variational wave function in 
Section~\ref{sec:Variational}. This wave function not only replicates the exact 
susceptibility at low fields but also describes the off-center 
displacement of the probability maximum for intermediate and high electric 
fields. In Sections \ref{sec:stressEquivalence}-\ref{sec:spectrumNarrowing}, we 
introduce valley-orbit and external stress effects, and in 
Section~\ref{sec:Zeeman-SO}, we describe the spin-orbit and Zeeman 
Hamiltonians. In Section~\ref{sec:GfacControl}, based on the results of the 
previous sections, we calculate the electric-field-induced ESR $g$-factor shifts for various types of 
spin-flip transitions and analyze the 
effects of random strains on the Stark shifts of the ESR spectra. 
Section~\ref{sec:Conclusions} contains the summary and conclusions.

\section{Quadratic Stark Effect}
\label{sec:Quadratic}

As a first step let us consider a quadratic Stark effect for a single-valley 
donor. We start with the EMT Kohn-Luttinger Hamiltonian~\cite{Kohn:1955vn} 
perturbed by an external electric field $\bf{\cal E}$ 
\begin{equation}
H_z = -\frac{\hbar^2}{2 m_\perp}\left(\spwrt{}{x}+\spwrt{}{y}+\gamma\spwrt{}{z}\right) 
- \frac{e^2}{\kappa r} - e\bm{\mathcal{E}}\cdot\bm{r},
\label{eqn:HKL}
\end{equation}
where $\kappa$ is the dielectric constant, 
$\gamma=m_\bot/m_\parallel$ is the effective mass anisotropy parameter, 
$m_\bot$ and $m_\parallel$ are transverse and longitudinal effective masses 
respectively, and we assume that the heavy-mass axis of the valley is along 
$z$.
For our purposes it is convenient to rewrite  the Hamiltonian  \eqref{eqn:HKL} using a scaling transformation 
$z\rightarrow\sqrt{\gamma} z$. This yields
\begin{equation}
H_{z\rho} = 
-\frac{\hbar^2}{2m_\perp}\left(\spwrt{}{x} + \spwrt{}{y} + \spwrt{}{z}\right) -\frac{e^2}{\kappa\rho} - e\bm{\mathcal{E}}\cdot\bm{\rho},
\label{eqn:scaledHKL}
\end{equation}
where $\bm{\rho}=(x,y,\sqrt{\gamma}z)$.

Our immediate goal is to find a second order shift of the ground state energy 
produced by an external static electric field $\bm{\mathcal E}$ (Stark shift):
\begin{align}
\delta_{z}(\mathcal{E}) 
&=\sum_{n\ne 1s} \frac{\langle 1s |e\bm{\mathcal{E}}\cdot{\bm\rho}|n\rangle\langle n| e\bm{\mathcal{E}}\cdot{\bm\rho}|1s\rangle}{E_{1s}-E_n} \nonumber \\
&=-\frac{1}{2}\chi_{\alpha\beta}{\cal E}_\alpha{\cal E}_\beta, 
\label{eqn:2nd-order}
\end{align} 
where we introduced the Stark susceptibility tensor and assumed summation over repeating Greek indices. In the
chosen coordinate system, the tensor  $\chi_{\alpha\beta}$ is diagonal and has only  two distinct
components $\chi_{zz}=\chi_\parallel$ and $\chi_{xx}=\chi_{yy}=\chi_\bot$ due to the axial symmetry of the
valley with
respect to  $z$.

It is well known from the classical problem of the quadratic Stark effect in
a hydrogen atom that any perturbative treatment of the hydrogen 
Schr\"{o}dinger equation in an external electric field requires summation of 
infinite series to account for the excited states of the continuous 
spectrum.\cite{Merzbacher:1998vw,Privman:1980wl,Alvarez:1991vp,Ivanov:1997uy} 
To accomplish this for the lithium donor, we will utilize the Dalgarno-Lewis exact summation method 
\cite{Dalgarno:1955bi}. We define a vector function ${\bm f}(\bfr)$ to satsify 
the equation
\begin{equation}
 \left[{\bm f}(\bfr), H\right]\ket{1s} = - {\bm\rho}\ket{1s}.
\label{eqn:def-of-f}
\end{equation}
With this equation in hand, consider an important condition for computing the 
second-order summation:
\begin{align}
\bracket{n}{\bm\rho}{1s} &= - \bracket{n}{\left[{\bm f},H\right]}{1s} \nonumber\\
                         &= (E_n - E_{1s})\bracket{n}{\bm f}{1s}. 
\end{align}
As a consequence, if the solution $\bm{f}(\bfr)$ of Eq.~(\ref{eqn:def-of-f}) 
is known, the Stark susceptibility tensor defined in Eq.~(\ref{eqn:2nd-order}) 
can be obtained as
\begin{equation}
\chi_{\alpha\beta}=2e^2\langle 1s|r_\alpha\cdot f_\beta(\bfr)|1s\rangle.
\label{eqn:chi}
\end{equation}

As both, the potential terms of $H$, as well as ${\bm f}(\bfr)$ are functions of the 
coordinates only, they will commute and thus $\left[{\bm f},H\right] = 
-(\hbar^2/2m_\bot)\left[{\bm f}, {\nabla}^2\right]$.
Expressed in the coordinate-representation, Eq.~(\ref{eqn:def-of-f}) becomes a 
differential equation defining the function $\bm{f}(\bfr)$ such that 
\begin{equation}
\nabla^2 {\bm f}(\bfr) + 2(\bm{\xi}\cdot{\bm \nabla}) {\bm f}(\bfr) 
= -\frac{2m_\bot}{\hbar^2}\bm{\rho},
\label{eqn:diffCond}
\end{equation}
where $\bm{\xi}={\bm\nabla}\ln(\psi_{1s})$, and $\psi_{1s}(\bfr)\equiv
\bra{\bfr}1s\rangle$ is the ground state wave function. 

To describe the ground state 
wave function $\psi_{1s}(\bfr)$ we use the Kohn-Luttinger function~\cite{Kohn:1955vn,Faulkner:1969uj} with 
ellipsoidal symmetry,
\begin{equation}
	\psi_{1s}(\bfr) 
		= \frac{\beta^{1/4}}{\left(\pi a_\bot^3\right)^{1/2}}
		  \exp\left(-\frac{\sqrt{x^2 + y^2 +\beta z^2}}{a_\bot} \right),
\label{eqn:variational}
\end{equation}
where $\beta = \gamma a_\perp^2/a_\parallel^2$ and $a_\perp, a_\parallel$ are 
the transverse and longitudinal radii of the isolated valley in the absence of 
the electric field. Then
\begin{equation}
\bm{\xi} = -\frac{1}{a_\perp}
\frac{x\hat{x}+y\hat{y} + \beta z\hat{z}}{\sqrt{x^2+y^2+\beta z^2}} .
\label{eqn:zeta}
\end{equation}

At this point it is convenient to introduce another scaling transformation 
such that $x\rightarrow a_\bot x$, $y\rightarrow a_\bot y$, and $ z 
\rightarrow a_\bot z/\sqrt{\beta}$. Under this transformation all the 
coordinates become dimensionless (measured in units of $a_\bot$) and the 
ground state wave function $\psi_{1s} \rightarrow \exp(-r)/\sqrt{\pi}$. 

Also, in Eq.~(\ref{eqn:diffCond}) we explicitly separate the anisotropic term proportional 
to $\lambda=1-\beta$ and treat it as a perturbation (i.e. we will construct our solution as a power series
in the anisotropy parameter $\lambda$). 
This yields:
\begin{equation}
\hat{D}_r \bm{f} - \lambda\hat{D}_z \bm{f} 
=- \frac{2m_\perp {a_\perp}^3}{\hbar^2}\bm{\zeta},
\label{eqn:separated}
\end{equation}
where $\bm{\zeta}=\left(x,y,z\sqrt{\gamma/\beta}\right)$, and
\begin{subequations}
\label{DrDz}
\begin{align}
\hat{D}_r&=\nabla^2 - 2{\partial}/{\partial r},\\
\hat{D}_z&={\partial^2}/{\partial z^2} - 2({z}/{r}){\partial}/{\partial z}.
\end{align}
\end{subequations}

From the axial symmetry of Eq.~(\ref{eqn:separated}), we are looking for
solutions in the form:
\begin{subequations}
\label{eqn:fxyz-ansatz}
\begin{align}
f_x(\bm{r})&=
(a_\perp/E_\perp)\cos\phi \cdot f_\bot(r,\theta)
\label{eqn:fx},\\
f_y(\bm{r})&=
(a_\perp/E_\perp)\sin\phi \cdot f_\bot(r,\theta)
\label{eqn:fy},\\
f_z(\bm{r})&=
(a_\parallel/E_\perp)\cdot f_\parallel(r,\theta)
\label{eqn:fz},
\end{align}
\end{subequations}
where $E_\perp=\hbar^2/2m_\perp a_\perp^2$ and $a_\|=a_\perp\sqrt{\gamma/
\beta}$. The partial differential equation \eqref{eqn:separated} may be 
further simplified by generating a coupled system of ordinary 
differential equations. The explicit angular dependence of the operators 
$\hat D_r$ and $\hat D_z$ is presented and analyzed in Appendix 
\ref{Dalgarno}. The analysis suggests that $f_\|(r,\theta)$ and $f_\bot(r,\theta)$ can be expanded into series in Legendre's and associated Legendre's polynomials, respectively, as follows:
\begin{subequations}
\label{eqn:Legendre-expansion}
\begin{align}
f_\parallel(r,\theta) &= \sum_{l=1} f_{\parallel,l}(r)P_l(\cos\theta)\label{eqn:Pl},\\ 
f_\bot(r,\theta) &=-\sum_{l=1} f_{\bot,l}(r)P^1_l(\cos\theta)\label{eqn:Pl1}.
\end{align}
\end{subequations}
Note that the series expansion begins at $l = 1$.

Substituting the Legendre's expansions \eqref{eqn:Pl} and 
\eqref{eqn:Pl1} into Eqs~\eqref{eqn:fx}-\eqref{eqn:fz}, inserting 
the results into Eq.~\eqref{eqn:chi} for the Stark susceptibility tensor, and integrating over the angular 
variables we obtain:
\begin{subequations}
\begin{align}
\chi_{\parallel}  = 
&= \frac{8 e^2 a_\|^2 }{3E_\perp} \int_0^\infty r^3 \exp(-2r) f_{\parallel,1}(r)\,\diff r,\\
\chi_{\perp}      = 
&= \frac{8 e^2 a_\perp^2 }{3E_\perp} \int_0^\infty r^3 \exp(-2r) f_{\bot,1}(r)\,\diff r,
\end{align}
\label{eq:chiInt}
\end{subequations}

The final task is to compute the radial parts $f_{\perp,l}(r)$ and 
$f_{\parallel,l}(r)$ of our function $\bm{f}$.
We accomplish this by inserting the combination of 
Eqs~\eqref{eqn:fx}-\eqref{eqn:fz} and \eqref{eqn:Pl}-\eqref{eqn:Pl1} into 
Eq.~\eqref{eqn:separated}. This procedure results in the following system of 
ordinary differential equations for the radial functions defined when $l>0$ 
(see Appendix~\ref{Dalgarno} for more details):
\begin{widetext}
\begin{subequations}
\label{eqn:chain}
\begin{align}
r^2\hat{D}_r f_{\parallel,l} - l(l+1)f_{\parallel,l} &= 
\lambda(\hat{\alpha}^0_{l}f_{\parallel,l+2} + \hat{\beta}^0_lf_{\parallel,l} + \hat{\gamma}^0_{l}f_{\parallel,l-2}) -r^3\delta_{l1},
\\
r^2\hat{D}_r f_{\bot,l} - l(l+1)f_{\bot,l} &= 
\lambda(\hat{\alpha}^1_{l}f_{\bot,l+2} + \hat{\beta}^1_lf_{\bot,l} + \hat{\gamma}^1_{l}f_{\bot,l-2}) -r^3\delta_{l1},
\end{align}
\end{subequations}
where $\hat{\alpha}_l^m, \hat{\beta}_l^m, \hat{\gamma}_l^m$ are second order 
radial differential operators having the common structure
\begin{equation}
\label{eqn:alpha-operators}
\hat{\alpha}^m_l \!=\!
\alpha^{(0)}_{l,m}\left[r^2\frac{\partial^2}{\partial r^2}
\!+\!\left(\alpha^{(1)}_{l,m} -2r\right)r\frac{\partial}{\partial r}
\!+\alpha^{(2)}_{l,m}r+\!\alpha^{(3)}_{l,m}\right],
\end{equation}
\end{widetext}
with different coefficients $\alpha^{(i)}_{l,m}$, $\beta^{(i)}_{l,m}$, and 
$\gamma^{(i)}_{l,m}$ as specified in Appendix~\ref{Dalgarno}. Note that 
$\hat{\gamma}^0_1 = \hat{\gamma}^1_1 = 0$. 

The solution to Eq.~\eqref{eqn:chain} with $\lambda = 0$ is known (see 
Appendix~\ref{Dalgarno}). This suggests that a perturbative treatment of the 
differential problem is appropriate. With this in mind, we expand 
$f_{\perp,l}(r)$ and $f_{\parallel,l}(r)$ as power series in $\lambda=1-
\beta$. Letting $q = \{\parallel\,,\,\perp\}$, our radial functions take the form:
\begin{align}
f_{q,l}(r)&=f_{q,l}^{(0)}(r) + \lambda f_{q,l}^{(1)}(r) + \lambda^2 
f_{q,l}^{(2)}(r) + \ldots 
\end{align}
Equating terms of like order in $\lambda$,
we obtain a chain of differential equations for different $l \ge 1$ and 
$n\ge 0$:
\begin{widetext}
\begin{subequations}
\begin{align}
r^2\hat{D}_r f_{\parallel,l}^{(n)} - l(l+1)f_{\parallel,l}^{(n)} &= (1-\delta_{n0})\left(
\hat{\alpha}^0_{l}f_{\parallel,l+2}^{(n-1)} + \hat{\beta}^0_lf_{\parallel,l}
^{(n-1)} + \hat{\gamma}^0_{l}f_{\parallel,l-2}^{(n-1)}\right) - r^3\delta_{l1}\delta_{n0} \\
r^2\hat{D}_r f_{\perp,l}^{(n)} - l(l+1)f_{\perp,l}^{(n)} &=(1-\delta_{n0})\left( 
\hat{\alpha}^1_{l}f_{\perp,l+2}^{(n-1)} + \hat{\beta}^1_lf_{\perp,l}^{(n-1)} 
+ \hat{\gamma}^1_{l}f_{\perp,l-2}^{(n-1)}\right) - r^3\delta_{l1}\delta_{n0}
\end{align}
\label{eq:iterative}
\end{subequations}
\end{widetext}

\begin{figure}[tbph]
{
	\includegraphics[width=3.0in]{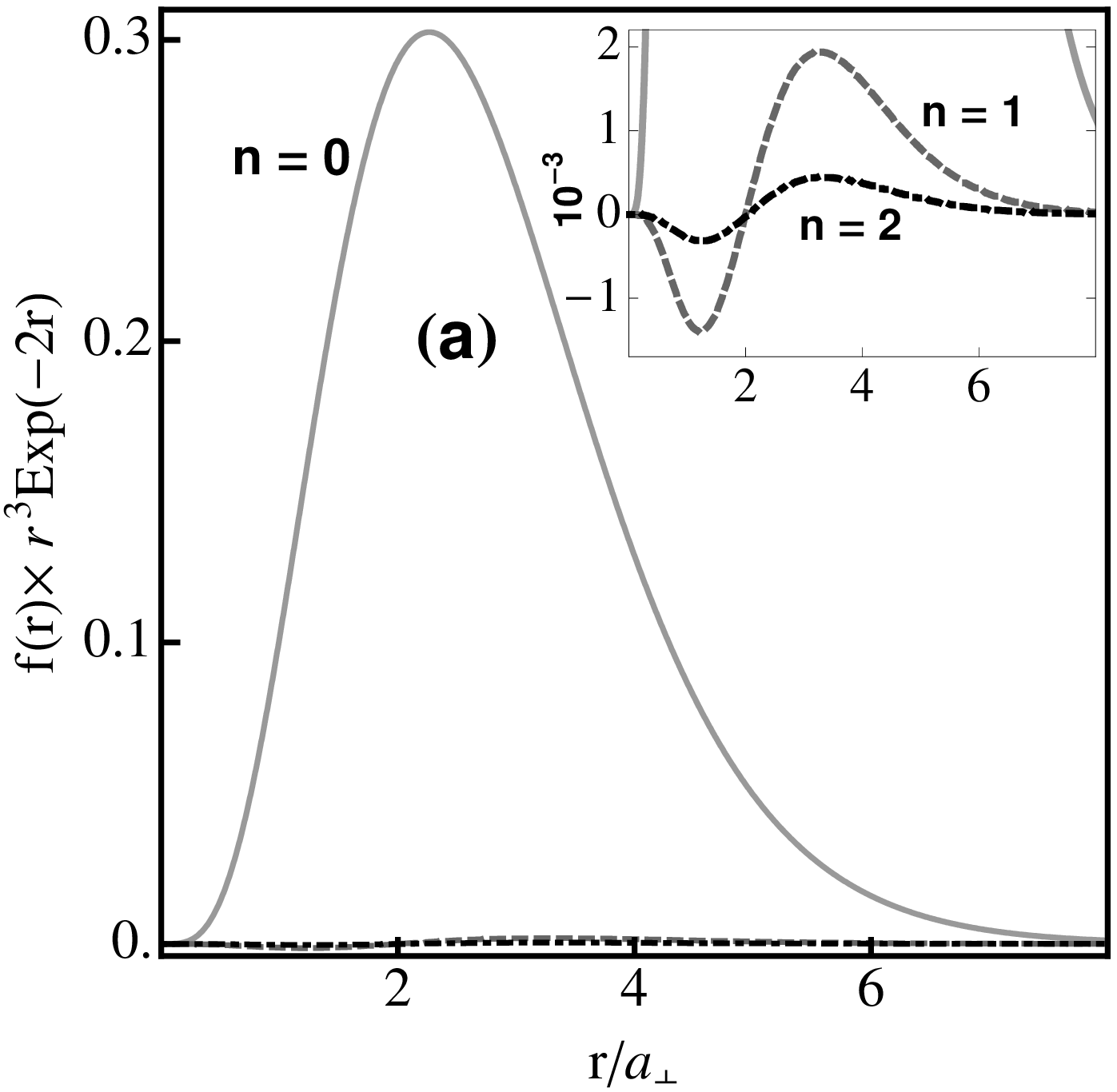}
	\label{fig:gxplot}
}
{
	\includegraphics[width=3.0in]{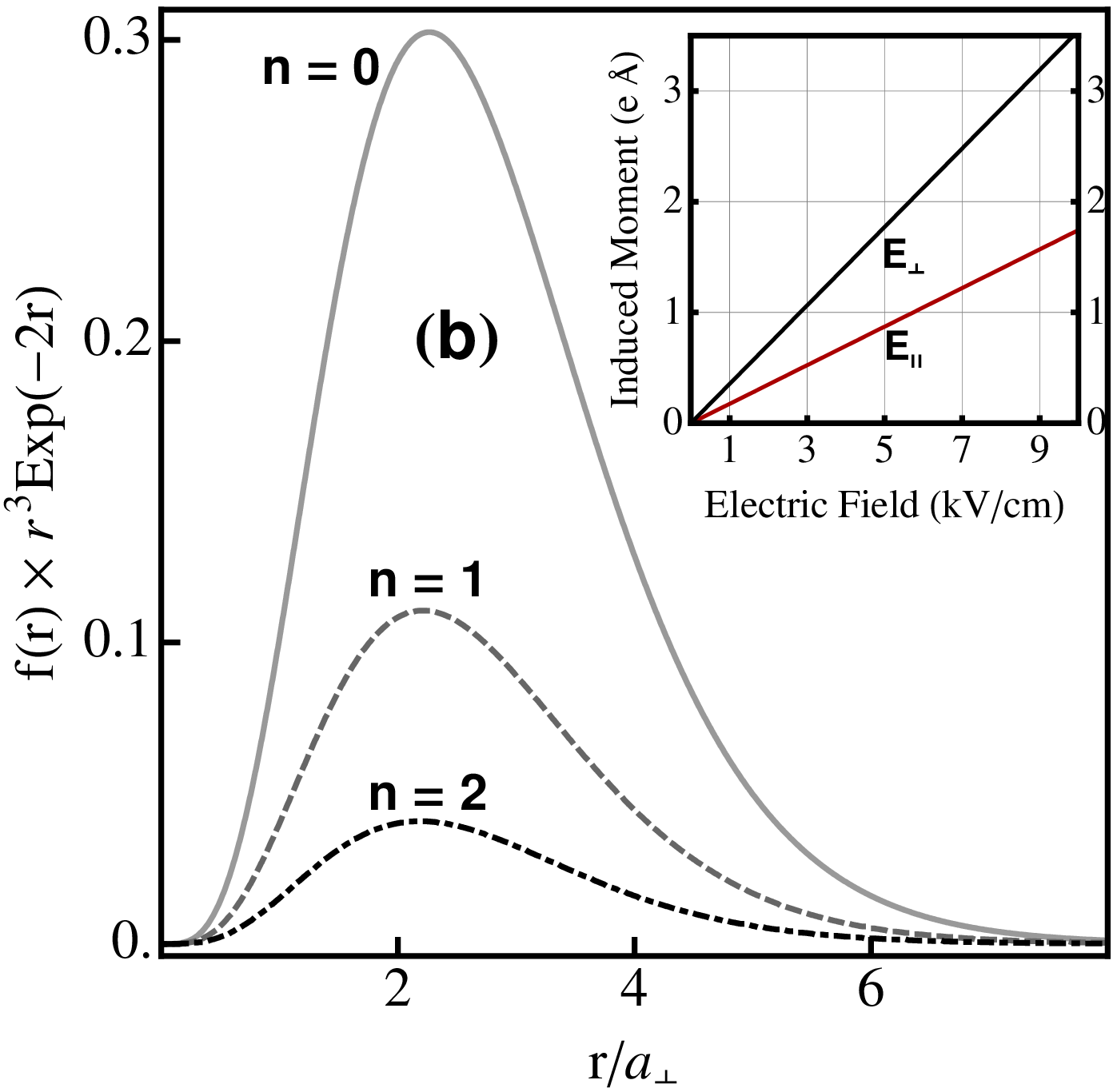}
	\label{fig:gzplot}
}
\caption{(Color online.)
The 
figures show successive integrands of (\ref{eq:chiInt}) up to the second order 
term for fields lying in the $x$-direction (a) and the $z$-direction (b). The vertical axes are scaled appropriately for each graph, and the legend for 
both figures is shown in (b). The inset in (a) is a magnified view of the two dashed lines
($n=1,2$) that are not resolved in the main plot. The inset in (b) displays dipole moments induced by the electric
fields parallel and perpendicular to the heavy-mass axis $z$.
}
\label{fig:gplot}
\end{figure}

From this point forward, the solution method will be entirely iterative. 
Determination of $f^{(1)}_{q,l}$ requires $f^{(0)}_{q,l}$, $f^{(2)}_{q,l}$ 
requires $f^{(1)}_{q,1}$ and so forth. Note that Eqs.~\eqref{eq:iterative} 
with $n = 0$ reduce to Eqs.~\eqref{eqn:chain} with $\lambda=0$. Since we know 
the solution for $\lambda=0$ (see Appendix~\ref{Dalgarno}), we use it as zeroth 
iteration, solve each differential equation for $f_{q,l}^{(1)}$ and proceed to the higher orders. The first few 
solutions are:
\begin{subequations}
\begin{align}
f_{\parallel,l}^{(0)} &= f_{\perp,l}^{(0)} = \frac{r(r+2)}{4}\delta_{l1}, \\
f_{\parallel,1}^{(1)} &= \frac{r(r+3)}{5}, \\
f_{\perp,1}^{(1)} &= \frac{r(r-2)}{40}.
\end{align}
\end{subequations}
The analytic form of the solutions past this point quickly becomes cumbersome. 
However, graphing the integrands $\lambda ^n r^3 \exp(-2r)f_{q,1}^{(n)}(r)$ of 
the first three terms shows the rapid convergence of the series, as seen in 
Fig.~\ref{fig:gplot}.

The final expressions for the susceptibilities, utilizing the expanded $f$, 
\begin{subequations}
\begin{align}
\chi_\parallel &= \frac{8 e^2 a_\|^2 }{3E_\perp} 
(0.84375 + 0.8250\lambda + 0.8173\lambda^2 + \ldots), 
\\
\chi_\perp &= \frac{8 e^2 a_\bot^2 }{3E_\perp}
(0.84375 + 0.0094\lambda + 0.0058\lambda^2 + \ldots), 
\end{align}
\end{subequations}
where for silicon $m_\perp = .191m_e$,
$m_\parallel = 
.916m_e$, $a_\perp=23.65$ ~\AA, and 
$a_\parallel = 13.60$~\AA.  For these values of the effective masses and Bohr radii $E_\perp = 35.68\,
\mathrm{meV}$, $\lambda = 0.370$, and our susceptibilities are 
\begin{subequations}
\begin{align}
\chi_\parallel &\approx 1.74 \mathrm{\mu eV (kV/cm)^{-2}}\label{chi_pert_para}
\\
\chi_\perp &\approx 3.54 \mathrm{\mu eV (kV/cm)^{-2}}.\label{chi_pert_perp}
\end{align}
\end{subequations}

\section{Variational Method}
\label{sec:Variational}

The findings of Section~\ref{sec:Quadratic} provide very accurate asymptotic behavior
of the energy levels at low electric fields. To achieve high accuracy for
the Stark shifts of the ESR lines both at low and intermediate fields $\sim$5 kV/cm  we have to extend
the scope of our methodology and use a specially crafted variational approach that is  valid
at higher fields and replicates the low-field results of  the previous section.

Thus 
we seek to devise 
and use a trial function which will adequately represent the perturbed 
donor state, accurate up to at least the second order. This variational 
function will be used to extract expansions of the energy and the central cell 
electron density, which cannot be obtained from the second order perturbation theory.  
At low fields we will see converging results between the variational method of
this section  and the infinite-series perturbation theory of Section~\ref{sec:Quadratic}.

In EMT, we can represent the orbital states of donors in silicon with 
hydrogenic-like envelope functions. We expect that a homogeneous electric field 
will admix $p$-state components into this ground state. In hydrogen, each $p$-function with 
principle quantum number $n$ will be of the form 
$rL_{n-2}^3(r/n)\exp(-2r/n)$ where $L_{n-2}^3$ is the generalized Laguerre 
polynomial of degree $n-2$. The higher $p$-states will contribute terms of 
order $r^{n-1}$ to the perturbed state.

The higher $p$-states with principle quantum numbers $n\ge 3$ couple somewhat weaker to 
the ground state. However, since there are many of these states, it is reasonable to surmise
that any highly accurate approximate method must account for their contributions. In summary, to 
accurately reflect the influence of the electric field on the hydrogenic 
wave functions of donors, our trial function must take into 
account the higher order radial contributions from the excited states.

It is instructive to illustrate our approach by considering the standard Hamiltonian for a hydrogen atom placed in a homogeneous electric field 
parallel to $z$-axis:
\begin{equation}
H = -\frac{\hbar^2}{2m_e}\nabla^2 - \frac{e^2}{r} - 
e\mathcal{E}z.
\end{equation}
It can be shown that the exact first order expansion of the 
wave function, and the associated second order expansion of the ground state energy are~\cite{Merzbacher:1998vw,Privman:1980wl}
\begin{align}
&\psi(\bm{r}) = \psi_0(\bm{r}) + e\mathcal{E} f(\bm{r})
\psi_0(\bm{r}), 
\label{psi1}\\
&E 
= -E_{Ry}-e^2\mathcal{E}^2\bra{\psi_0}z f(\bm{r})\ket{\psi_0},
\label{E2}
\end{align}
where $f(\bm{r}) = (1/a_BE_{Ry})(r+2a_B)z/4$ with $a_B$ and $E_{Ry}$ 
are the Bohr radius and the Rydberg energy respectively. The function $f(\bm{r})=f_j(\bm{r})$ with
$j=x,y,z$,
is the 
isotropic analog  of the function $\bm{f}(\bm{r})$ considered in the infinite-series summation of
the previous section.
The wave function \eqref{psi1} can be recast as
\begin{equation}
\label{eq:1st-order-wf}
\psi(\bm{r}) = \left[1 + \left(q_1 + q_2r\right)z\right]\exp(-r/a),
\end{equation}
where $q_1 = e\mathcal{E}/2E_{Ry}$, $q_2 = e\mathcal{E}/4a_BE_{Ry}$, and $a=a_B$.

To extend our formalism to higher orders we can identify $q_1$
and $q_2$ as the first order expansion of some unknown {\em variational} 
parameters. Therefore we assume that the trial function takes the form of Eq.~\eqref{eq:1st-order-wf} 
with the unknown variational parameters $q_1=q_1(\mathcal{E})$, $q_2=q_2(\mathcal{E})$, and $a=a(\mathcal{E})$ to be determined via a standard minimization routine at an arbitrary field $\mathcal{E}$. This procedure leads 
to the following expansion of the energy expectation value: 
\begin{equation}
\left\langle E \right\rangle 
= \frac{\bra{\psi}H\ket{\psi}}{\braket{\psi}{\psi}} 
= -E_{Ry} - \frac{9}{8}a_B^2e^2\mathcal{E}^2/E_{Ry} + \ldots,
\label{eq:exact-E}
\end{equation}
which is in complete agreement with the exact second-order $\mathcal{E}$-field expansion of the
ground state energy \eqref{E2}.

The term proportional to $rz$ in Eq.~\eqref{eq:1st-order-wf} is responsible 
for admixture of the excited states 
into the ground state by the electric field. By including this term in
the trial function we were able to replicate the
exact second order correction to the energy.
In addition, 
the variational method allows for an efficient (albeit approximate) account of higher order 
terms outside of the 
perturbative regime. 
For the hydrogen atom,  
setting $q_2 = 0$ in Eq.~\eqref{eq:1st-order-wf} 
yields the following Taylor 
expansion of the expectation value of energy:
\begin{equation}
\left\langle E \right\rangle 
= -E_{Ry} - a_B^2e^2\mathcal{E}^2/E_{Ry} + \ldots \hskip 0.5 cm .
\label{eq:Friesen-E}
\end{equation}
Thus by neglecting contributions of the excited states, we would introduce 
a relative error of $11 \%$  to the second order energy shift, running the
risk of producing more significant errors in the higher orders. 

In the case of the silicon donor states, our variational wave function must 
be generalized to reflect the anisotropy of the effective mass. To take this 
anisotropy into account, we will construct the trial function in the form similar to the 
hydrogenic function of Eq.~\eqref{eq:1st-order-wf}, but with the appropriately scaled
exponential and pre-exponential factors. Due to the axial symmetry
of the Hamiltonian \eqref{eqn:HKL} at $\mathcal{E}=0$  the $\mathcal{E}$-field vector 
can be chosen to lie in  the ($xz$)-plane  without any loss of generality.
Thus our final 
variational function reads:
\begin{equation}
F_{+z}(\bm{r}) =F_z(\mathcal{E}) \left(1 + Q_{\perp}x + Q_{\parallel}z\right)
\exp(-\varrho/a_\perp)
\label{my_var_func},
\end{equation}
where $Q_\perp = q_{1\perp} + q_{2\perp}\varrho$, $Q_\parallel= q_{1\parallel} + 
q_{2\parallel}\varrho$ , $\varrho = \sqrt{r^2 + (a_\perp^2/a_\parallel^2-1)z^2}$, and
$F_z(\mathcal{E})$ is the normalization constant. Similar variational functions with normalization
constants $F_j(\mathcal{E})$ can be defined for any valley $sj$.

Friesen investigated the Stark effect for a phosphorous 
donors in silicon~\cite{Friesen:2005fc} using a single-valley variational method and a degenerate perturbation theory 
(see Eq.~\eqref{eq:Hvo} below) to account for the valley-orbit effects. 
Our trial function~\eqref{my_var_func} will reduce to that of Friesen's if we force
parameters $q_2$ to zero, i.e. set and fix $q_{2\perp}=q_{2\parallel}=0$. As we have seen 
from Eq. \eqref{eq:Friesen-E}  it may lead to some inaccuracy in the low-filed limit.
In what follows we will use our improved single-valley variational function \eqref{my_var_func}
and adopt Friesen's treatment of the valley-orbit effects.
First, we compute the expectation energy of the single valley states, 
neglecting central cell contributions. For given $\gamma$ and $\mathcal{E}$ we 
take the expectation value of energy $E_{sj}$ to be a function of the parameters $a_\eta$, and 
the various $q_{1\eta}$ and $q_{2\eta}$ with $\eta = \{\parallel, \perp\}$. The energy
\begin{equation}
E_{sj}(a_\eta,q_{1\eta},q_{2\eta};\gamma,\mathcal{E}) = \frac{\bra{F_{sj}}H_{sj}\ket{F_{sj}}}{\braket{F_{sj}}{F_{sj}}},
\end{equation}
is minimized with respect to these parameters:
 \begin{gather}
\pwrt{E(\gamma,\mathcal{E})}{a_\eta} = \pwrt{E(\gamma,\mathcal{E})}{q_{1\eta}} = \pwrt{E(\gamma,\mathcal{E})}{q_{2\eta}} = 0.
\label{eq:minimization}
\end{gather}
This condition implicitly defines the variational parameters 
as functions of the electric field. Using this knowledge, we can expand the  single-valley ground state energy  in Taylor series around zero field. 

A simple system to test the variational function~\eqref{my_var_func}, is a hypothetical
hydrogenic donor with the isotropic effective mass, $m_\perp=m_\parallel$, 
i.e. $\gamma=1$. Since the unperturbed 
Hamiltonian displays spherical symmetry, we can take the $z$-axis to lie along 
the field direction. If we define our atomic units of energy and distance as
$E_0 = \hbar^2/2m_\perp a_0^2$ and $a_0 = \hbar^2\kappa/2m_\perp e^2$, respectively, and  minimize
the energy according to \eqref{eq:minimization} we obtain in the low-field limit: $q_{1\perp}=q_{2\perp}=0$, 
$q_{1\parallel}
= e\mathcal{E}/2E_{0}$, and $q_{2\parallel}= e\mathcal{E}/4a_0E_0$, 
in other words, we recover the exact expansions \eqref{eq:1st-order-wf}-\eqref{eq:exact-E}, where we replace
$a_B\rightarrow a_0$ and $E_{Ry}\rightarrow E_0$. Similarly for the Friesen
case with $q_{2\parallel}=0$ we obtain a less accurate expansion \eqref{eq:Friesen-E},
which does not comply 
with Rayleigh-Schr\"{o}dinger perturbation theory.

It is expected that our variational function will return better corrections 
than previous methods, especially when used for small fields. 
For silicon, the effective mass anisotropy parameter $\gamma =0.209$. We 
can no longer freely rotate the coordinate system due to a fixed heavy mass 
axis, and we will be required to keep track of the components of electric 
field parallel and perpendicular to this axis. Consider an envelope
function of the valley $sj$, $F_{sj}(\bm{r},\bm{\mathcal{E}})$.
To study the valley-orbit corrections due to the central cell contact potential 
we will need the values of $F_{sj}(0,\bm{\mathcal{E}})\equiv F_{j}(\mathcal{E})$ 
at ${\bm r} = 0$. Our computations proceed as in the isotropic hydrogenic case, and we obtain 
the second-order expansion of the energy and the central cell 
amplitudes: 
\begin{subequations}
\begin{align}
E &= E_{0} - \frac{1}{2}\chi_{\parallel}\mathcal{E}_\parallel^2 
- \frac{1}{2}\chi_{\perp}\mathcal{E}_\perp^2, \\
F_{j}(\bm{\mathcal{E}})/F_0 &= 1 - \frac{1}{2}f^{(2)}_{\parallel}\mathcal{E}_\parallel^2 - \frac{1}{2}f^{(2)}_{\perp}\mathcal{E}_\perp^2.
\end{align}
\end{subequations}
Here, $\mathcal{E}_\parallel$ and $\mathcal{E}_\perp$ are the field components parallel and perpendicular to
the heavy-mass axis of the valley $sj$, $F_0=F_j(0)$ is the normalization constant at zero field,
and the numerical values of the susceptibilities and coefficients $f^{(2)}$
are:
\begin{subequations}
\begin{align}
\chi_{\parallel} 
&= 1.71 \,\mathrm{\mu eV(kV/cm)^{-2}}, \\
\chi_{\perp} 
&= 3.63 \,\mathrm{\mu eV(kV/cm)^{-2}}, \\
f^{(2)}_{\parallel} 
&= 1.53\times 10^{-4} \,\mathrm{(kV/cm)^{-2}}, \\
f^{(2)}_{\perp} 
&= 2.82\times 10^{-4} \,\mathrm{(kV/cm)^{-2}}.
\end{align}
\end{subequations}
We see good agreement with the susceptibilities presented in 
Eqs.~(\ref{chi_pert_para}) and (\ref{chi_pert_perp}). Additionally, we can 
check the answers obtained using Friesen's previous results. As before, by 
setting $q_{2\perp}$ and $q_{2\parallel}$ to zero and holding them fixed, we 
obtain equivalent results for Friesen's variational function:
\begin{subequations}
\begin{align}
\chi_{\parallel} 
&= 1.58 \,\mathrm{\mu eV(kV/cm)^{-2}}, \\
\chi_{\perp} 
&= 3.17 \,\mathrm{\mu eV(kV/cm)^{-2}}, \\
f^{(2)}_{\parallel} 
&= 1.54\times 10^{-4} \,\mathrm{(kV/cm)^{-2}}, \\ 
f^{(2)}_{\perp} 
&= 2.69\times 10^{-4} \,\mathrm{(kV/cm)^{-2}},
\end{align}
\end{subequations}

\begin{figure}[htb]
{
	\includegraphics[width=3.0in]{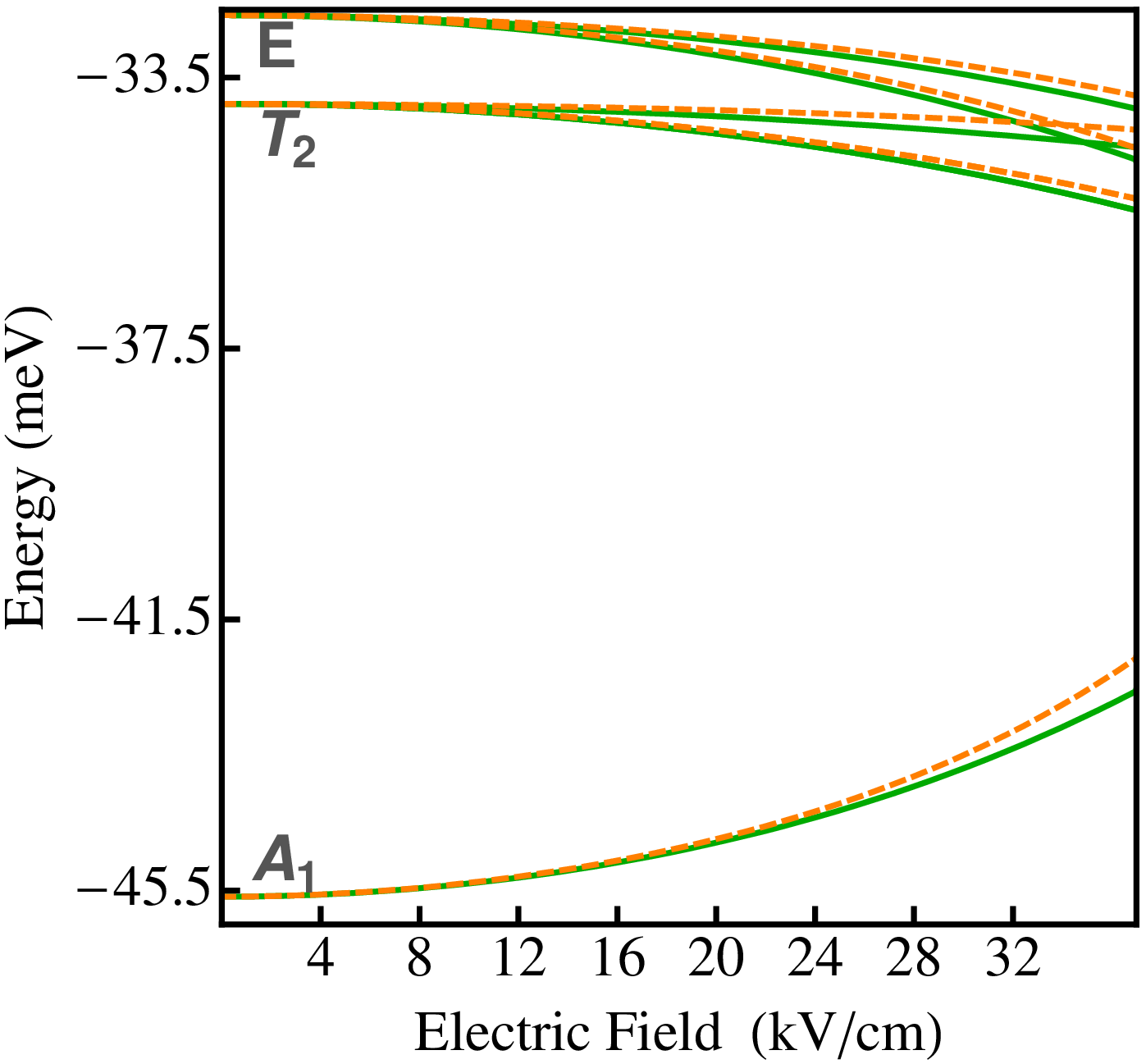}
	\label{fig:phosphorousPlt}
}
{
	\includegraphics[width=3.0in]{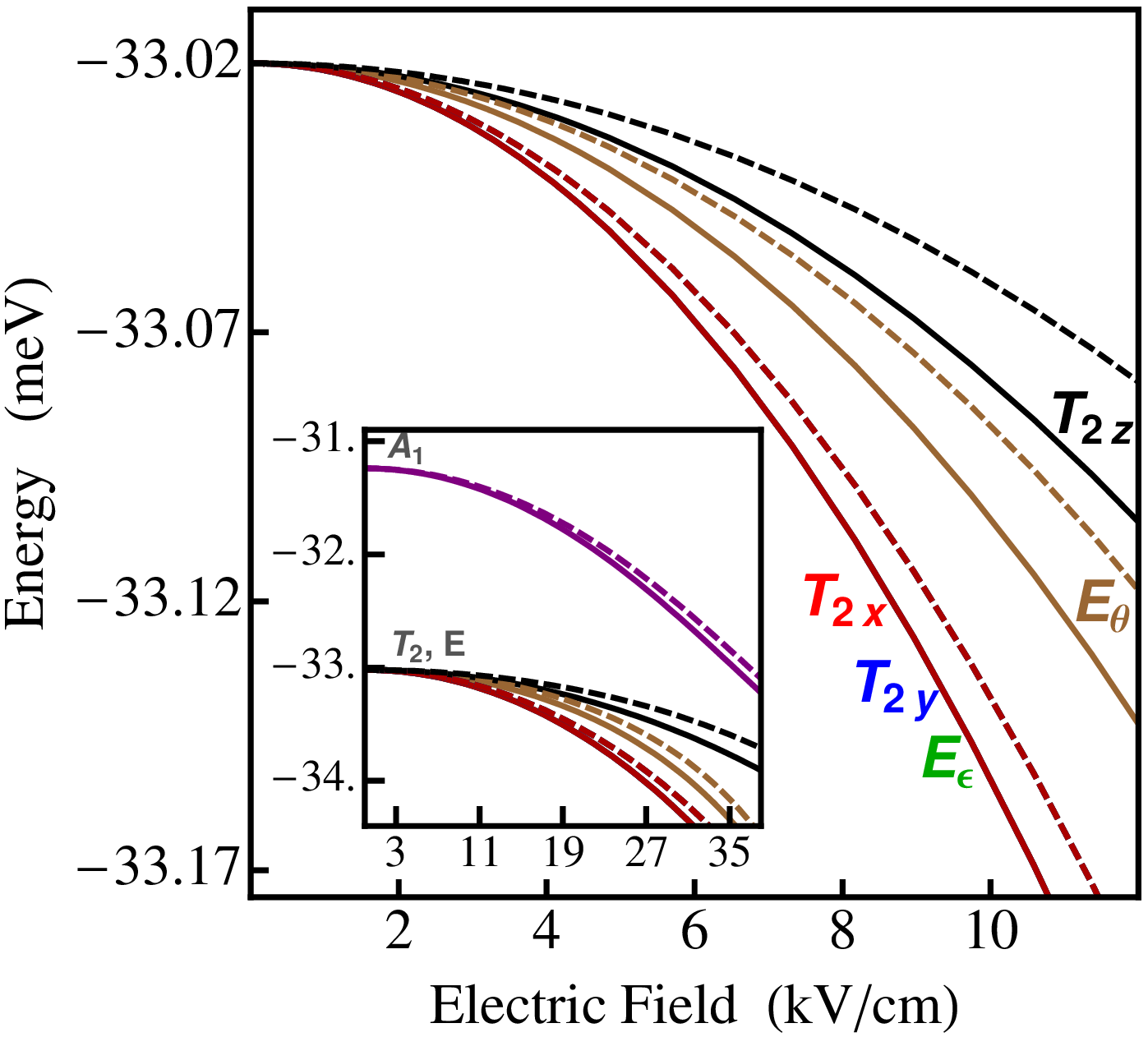}
	\label{fig:lithiumPlt}
}
\caption{
(Color online.) The plot on the top shows the comparison between our results (solid lines) 
and Friesen's (dashed) for the spectrum of a phosphorous donor in silicon. The second plot shows 
the spectrum of a lithium donor, where the dashed lines correspond to Friesen's 
variational function ($q_{2\perp}=q_{2\parallel}=0$) for Li.
}
\label{fig:StarkEffectPlt}
\end{figure}

With the variational functions at hand we can construct the valley-orbit 
Hamiltonian~\cite{Friesen:2005fc,Fritzsche:1962ww} in the presence of the electric field:
\begin{align}
H_{vo}(\bm{\mathcal{E}})&=
\sum_{s,i} \left(E_{0i}+\Delta_{0i}\right) \ket{si}\bra{si}
+\sum_{i,s}\Delta_{1i}\ket{si}\bra{ -s i}\nonumber\\
&{}+\sum_{i,j,s,s^\prime}\Delta_{2ij}
(1-\delta_{ij})\ket{si}\bra{s^\prime j},
\label{eq:Hvo}
\end{align}
where $E_{0i}$ may be written as $E_{0i} = E_0 - \frac{1}{2}\chi_{\perp}
\mathcal{E}^2 + \frac{1}{2}\left(\chi_\perp-\chi_\parallel\right)
\mathcal{E}_{i}^2 + ...$~.  Other corrections to the terms involving 
$E_{0i}$ will be left to the next section. The matrix elements
$\Delta_q$ of the  central cell contact potential between the valley-orbitals \eqref{valley-orbital}
describe the central cell shift ($\Delta_0$) and the valley-orbit splittings ($\Delta_1$ and $\Delta_2$)
of the energy levels.  Following 
Friesen's prescription,\cite{Friesen:2005fc} we parametrize the matrix 
elements $\Delta_q$ as follows:
\begin{subequations}
\begin{align}
&\Delta_{0j} = \nu_0 F_{j}^2(\mathcal{E}),\label{sim:delta0} \\
&\Delta_{1j} = \nu_1 F_{j}^2(\mathcal{E}),\label{sim:delta1} \\
&\Delta_{2ij} = \nu_2 F_{i}(\mathcal{E})F_{j}(\mathcal{E})\label{sim:delta2}.
\end{align}
\end{subequations}
The parametrization in 
Eqs.~\eqref{sim:delta0}-\eqref{sim:delta2} assumes that the central cell potential has 
a contact (i.e. $\delta$-function like) form. The fitting parameters $\nu_k$ depend on the choice of 
the variational functions to ensure that the experimental spectrum at zero 
field is reproduced correctly. For our variational functions defined by 
Eq.~\eqref{my_var_func}, it gives the following values of the parameters $\nu_i$ 
in Si:Li:
$\nu_0$=-34.54~eV$\cdot$\AA$^3$, $\nu_1$=7.09~eV$\cdot$\AA$^3$, and $\nu_2$=7.09~eV$\cdot$\AA$^3$.
In Fig.~\ref{fig:StarkEffectPlt}, we compare the spectra of P and Li shallow 
donors  calculated with ours and Friesen's variational functions for an 
external electric field in $001$ direction. For phosphorous impurity our approach does 
not produce substantial differences at low and intermediate fields. For lithium, on the other hand, 
the effect is more significant due to  the  nearly degenerate ground state. 

At this juncture, we need to consider the two competing effects caused by the electric field. 
The first effect is the direct quadratic Stark shift 
of the single-valley Coulomb binding energies $E_{0i}$. This effect will 
be similar to that of strain upon the donor spectrum, as we discuss below in 
section~\ref{sec:stressEquivalence}. 
The second effect is based on the fact that by 
pulling the lithium donor electron away from the central cell, the electric 
field reduces the magnitude of the matrix elements $\Delta_q$. 
This will bring the levels closer to their ``center of gravity"  and narrow the overall 
energy spectrum of the $1s$ manifold.~\cite{Friesen:2005fc} For Li donors subject to  relatively low electric
fields below ionization threshold, this {\em spectrum narrowing}  
effect 
may be treated separately and independently 
from the Stark shift of the 
valley energies $E_{0i}$ by means of the second-order expansion of the variational 
parameters, as detailed in Section~\ref{sec:spectrumNarrowing} and Appendix~\ref{narrowing}.

\section{Effect of Strain and Electric Field}
\label{sec:stressEquivalence}
We wish to consider the interplay of the Stark effect with the effects due to
other influences such as strain and magnetic field. Both the electric field 
and strain in the silicon lattice will cause a change  of the 
six single-valley energies that emerge on the diagonal of  the valley-orbit Hamiltonian~\eqref{eq:Hvo}. By virtue of the ellipsoidal valley symmetry these energies will not depend on
index $s$  and 
 for a given 
valley $si$, we can write both the quadratic Stark effect and the strain corrections to the 
energy as $E_{si}=E_{0i} = E_0 - \delta_i(\mathcal{E}) - \delta_i(e_{jk})$ such 
that 
\begin{subequations}
\begin{align}
\delta_i(\mathcal{E}) &= -\frac{1}{2}\chi_\perp\mathcal{E}^2 +
\frac{1}{2}\left(\chi_\perp-\chi_\parallel\right)\mathcal{E}_i^2, \label{quad-stark}\\
\delta_i(e_{jk}) &= \Xi_d(e_{xx}+e_{yy}+e_{zz})+\Xi_u e_{ii},\label{strain}
\end{align}
\end{subequations}
where $e_{jk}$ are the components of the strain tensor at the donor location and $\Xi_d$ and $\Xi_u$ are the 
dilation and shear 
deformation-potential constants of Si conduction band minimum.~\cite{Herring:1956vy}
The nearly degenerate ground state manifold of the Li donor 
is very sensitive to strain. However a simple dilation merely shifts the levels by the same amount, 
$\Delta E = (\Xi_d+\Xi_u/3)(e_{xx}+e_{yy}+e_{zz})$, and does not alter their separation. 
To change the relative positions of the energy levels and lift the fivefold degeneracy of the ground state, at least one of the   two  linear combinations of the strain tensor components 
\begin{equation}
e_\theta=e_{zz}-\frac{1}{2}(e_{xx}+e_{yy}),\quad e_\varepsilon=\frac{\sqrt{3}}{2}(e_{xx}-e_{yy})\label{eq:e}
\end{equation}
must be different from zero.~\cite{Watkins:1970wj}

In a similarity, a uniform electric field can also lift 
the degeneracy. The quadratic Stark effect can be described by the  variables
\begin{equation}
w_\theta=\frac{1}{2}\left(3n_{z}^{2}-1\right),\quad w_\epsilon=\frac{\sqrt{3}}{2}\left(n_{x}^{2}-n_{y}^{2}\right),\label{eq:w}
\end{equation}
where $n_{x, y, z}$ are components of the unit vector $\bm{n}$ in the direction 
of the electric field 
\begin{equation}
 \bm {\mathcal{E}}=\mathcal{E}\, \bm{n},\quad \bm{n}=(n_x,n_y,n_z).\label{eq:E}
\end{equation}

Therefor to account for the strain effects, 
we must add the strain energies \eqref{strain} to the diagonal  of the valley-orbit 
matrix  (see Eq.~\eqref{eq:Hvo}) in $|si\rangle$ basis.
To the second order in $\mathcal{E}$, we may combine the strain and Stark terms and separate them from the zero field valley-orbit 
Hamiltonian~\eqref{eq:Hvo0}. Fixing the average energy of the 
$1s$ manifold at zero, the combined Hamiltonian, describing strain and electric field 
effects, may be written as
\begin{equation}
\hat H_{S}= \Xi_u\left(v_\theta \, \hat V_\theta+v_\epsilon \, 
\hat V_\epsilon\right) ,\label{eq:Hss}
\end{equation}
where $\Xi_u = 11.4\,\mathrm{eV}$. 
Here $\hat{V}_\theta$ and $\hat{V}_\epsilon$ are operators in 
the space spanned by the six symmetrized orbitals $|\mu\rangle$
\begin{subequations}
\begin{align}
\hat V_\theta =&\frac{1}{3}\left(|E_\theta\rangle\langle E_\theta|-|E_\epsilon\rangle\langle E_\epsilon|-|T_{2x}\rangle\langle T_{2x}|-|T_{2y}\rangle\langle
T_{2y}| \right)\nonumber\\ 
&{}+\frac{2}{3}|T_{2z}\rangle\langle T_{2z}| + \frac{\sqrt{2}}{3}\left(|A_1\rangle\langle E_\theta|+|E_\theta\rangle\langle A_1 |\right), \label{eq:Vtheta} \\
\hat V_\epsilon =& \frac{1}{\sqrt{3}}\left(|T_{2x}\rangle\langle T_{2x}|-|T_{2y}\rangle\langle T_{2y}|\right)\nonumber\\
&{}+\frac{1}{3}\left(\sqrt{2}|A_1\rangle\langle E_\epsilon|-|E_\theta\rangle\langle E_\epsilon|+H.C. \right),\label{eq:Vepsilon}
\end{align}
\end{subequations}
where $H.C.$ denotes hermitian conjugation.

In Eq.~(\ref{eq:Hss}), the variables $v_\theta$ and $v_\epsilon$ are linear combinations of terms describing the effects of strain and electric field:
\begin{subequations}
\label{eq:effstrain}
\begin{eqnarray}
v_\theta &=& e_\theta+\kappa \,{\mathcal{E}}^2\,w_\theta,\label{eq:vtheta}\\
v_\epsilon &=&e_\epsilon+\kappa\, {\mathcal{E}}^2 \,w_\epsilon,\label{eq:vepsilon}
\end{eqnarray}
\end{subequations}
where 
\begin{equation}
\kappa=\frac{1}{2}\frac{\chi}{\Xi_u}\approx 10^{-7} \left(\frac{\rm kV}{\rm cm}\right)^{-2}, \label{eq:kappa}
\end{equation}
and $\chi=\chi_\perp - \chi_\parallel = 1.9\,\mathrm{\mu eV 
(kV/cm)^{-2}}$ is the anisotropic part of the quadratic Stark effect susceptibility. 
The quantities $v_\epsilon$ and $v_\theta$ in Eqs.~(\ref{eq:effstrain}) can be viewed
as ``effective strain" variables. The quadratic Stark effect due to the field 
$1\,\mathrm{kV/cm}$ is equivalent to a very small strain $\sim 10^{-7}$. 
Larger electric fields of the order of $3\,\mathrm{kV/cm}$ will be equivalent 
to the strain $\sim 10^{-6}$.

\begin{figure}[tb]
\includegraphics[width=3.0in]{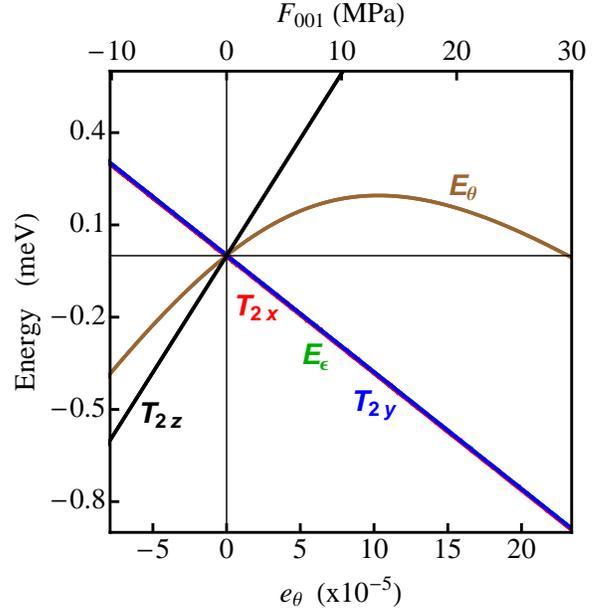}
\caption{(Color online.) Dependence of the 5 lowest orbital energy levels of Si:Li donor 
electron on the uniaxial [001] strain. Dominant characters of the states are 
given with letters. Color of the lines also encodes the dominant characters 
of the states for each value of $e_\theta$.}
\label{fig:uniaxial}
\end{figure}

Elaborating on the analogy between strain and electric field,  
the two effective strain parameters determine the orbital response 
when actual strain or electric fields are present. For 
simplicity, let us initially consider the case when $v_\epsilon = 0$ but 
$v_\theta$ is nonzero. As follows from Eqs.~\eqref{eq:e}, 
\eqref{eq:w}, \eqref{eq:vtheta} and \eqref{eq:vepsilon},
the degeneracy of the Li ground state quintet can be partially lifted
either with uniaxial strain along [100], or with biaxial isotropic strain 
in the ($xy$)-plane, or with some combination thereof.  The states
$T_{2x}$, $T_{2y}$ and 
$E_\epsilon$, which do not contain any contribution  from $|\pm z\rangle$,
valley-orbitals will remain degenerate. The nature of the strain determines the 
structure of the ground state. A tensile uniaxial strain along [001] has $v_\theta > 0$ and 
favors the ($T_{2x}$, $T_{2y}$, $E_\epsilon$) triplet ground state. 
Contrariwise, the non-degenerate ground state $T_{2z}$  and the 
first excited state  $E_{\theta}$ correspond to a compressive uniaxial [001] strain with $v_\theta < 0$.  These findings are summarized in Fig.~\ref{fig:uniaxial}.

One can relate the effective strains produced by an imposed external stress using the 
three-dimensional Hooke's law for an isotropic material: $e_\theta = (1+\nu)
(\sigma_{zz} - \frac{1}{2}(\sigma_{xx}+\sigma_{yy}))/E$ and $e_\epsilon = 
\frac{\sqrt{3}}{2}(1+\nu)(\sigma_{xx} - \sigma_{yy})/E$. Here,  $E$ and $\nu$
are the Young's modulus and Poisson's ratio, respectively,  and $\sigma_{jj}$ 
is the applied stress component along the $j$th axis. If we 
wish to create a condition with $e_\theta \ne 0$ and $e_\epsilon = 0$, we require $\sigma_{xx} = 
\sigma_{yy}$ and find $e_\theta = (1+\nu)(\sigma_{zz} - \sigma_{xx})
/E$. In short, to prepare a three-fold degenerate ground state ($T_{2x}$, $T_{2y}$, $E_\epsilon$), a uniaxial tension along [001] is sufficient. As we will see below, an additional stress along [100] or [010]
will  result in $v_\epsilon \ne 0$ and 
completely lift the orbital degeneracy.

We can create similar conditions with the electric field. To keep
$v_\epsilon = 0$ at nonzero field, the electric field must be confined to the 
plane formed by the $[001]$ axis and either the $[110]$ or the $[1\bar{1}0]$ 
axes. The uniaxial strain can be replicated by electric fields 
lying in this plane. An electric fields parallel to $[001]$  replicates 
tensile uniaxial strain, while fields parallel to either $[110]$ or $[1\bar{1}0]$ 
replicate compressive uniaxial strain. Pointing the  
electric field away from these axes, but still within the plane in question,  
only results in reducing the effective strain $v_\epsilon$, which vanishes  entirely for a field parallel to $[111]$. 

One can expect a non-trivial interplay of the Zeeman and Stark effects due to 
multiple level crossings in the ground state manifold. The possibility of the ground 
state Stark splitting makes the Li impurity unique among other shallow donors 
in Si and opens exciting opportunities for electrical manipulation of the 
spin qubits. As we will show below, the values of the electric field in 1-3 kV/cm 
range are sufficient to produce large changes in $g$-factors near the points of 
avoided crossing between the donor electron energy levels controlled by Zeeman 
and spin-orbit interaction.

To examine this interplay further, we want to introduce a second source of
effective strain $v_\epsilon$, in this case such that $|v_\theta| >> 
|v_\epsilon| > 0$. A nonzero strain $v_\epsilon$ separates the 
triplet manifold ($T_{2x}$, $T_{2y}$, $E_\epsilon$), allowing us to maximize 
the spin-orbit effects and control $g$-factors. A  biaxial strain anisotropy or an electric field 
lying in the ($xy$)-plane will produce nonzero $v_\epsilon$. 
In general, we should note that the increase of 
strain or electric field along a single crystallographic axis will change the 
value of $v_\theta$ by $v_\epsilon\sqrt{3}/3$. For our purposes, we consider 
$v_\theta \approx 10^{-4}$, which will not be seriously influenced by 
$v_\epsilon \approx 10^{-6}$. As long as $|v_\epsilon| << |v_\theta|$, the 
correction to $v_\theta$ will not introduce any serious errors to
the spin-orbit interactions within the ground state triplet 
($T_{2x}$, $T_{2y}$, $E_\epsilon$).

As before, we can think of the electric field as analogous to strain. In 
general, electric fields lying in the ($xy$)-plane will produce a nonzero 
effective strain $v_\epsilon$. The influence of the electric field on 
$v_\epsilon$ is maximal along the crystallographic axes and vanishes entirely 
along $[110]$ or $[1\bar{1}0]$.  The field along $[100]$ produces 
``tensile'' effective strain $v_\epsilon > 0$ while the field along $[010]$ produces a 
``compressive'' effective strain $v_\epsilon < 0$.

Despite having similar qualitative features, the electric field effects are 
considerably weaker then those of strain.  Additionally, strong electric 
fields will ionize the Li donor and may even induce electrical breakdown within the silicon. 
For this reason we will consider  only moderate electric fields in 1-3 kV/cm range
and envision their role as a means of precise fine tuning and control of the Li donor 
spectrum. We must rely on tensile strain $e_\theta$ to split the Li quintet and isolate 
the ($T_{2x}$, $T_{2y}$, $E_\epsilon$) ground state triplet.  The fine tuning by the electric field
will then induce 
much smaller effective strain $v_\epsilon$. By inducing the 
splitting of the ($T_{2x}$, $T_{2y}$, $E_\epsilon$) triplet, the electric 
field can be used to explore the effects of the spin-orbit interaction through 
$g$-factor control of ESR spectra.

\section{Effect of Spectrum Narrowing}
\label{sec:spectrumNarrowing}
Electric fields can influence the donor spectrum in a strain-like 
way, however, higher-order confounding effects emerge through the electric field 
dependence of the valley-orbit 
matrix elements $\Delta_{0j}$, $\Delta_{1j}$, and $\Delta_{2ij}$. 
Generally speaking, these spectrum narrowing effects violate  
relationship between strain and electric field. However,  the analogy described in Sec.~\ref{sec:stressEquivalence}, remains intact 
if we are only concerned with the influence of the electric field upon the isolated ($T_{2x}$, $T_{2y}$, 
$E_\epsilon$) manifold.

Appendix~\ref{narrowing} details a procedure for constructing the orbital Hamiltonian with
strain, electric field and spectral narrowing effects taken. 
Let us consider a truncated version of the 
Hamiltonian given in Eq.~\eqref{spectrum_narrowing} restricted to the 
$\{\ket{E_\epsilon}, \ket{T_{2x}}, \ket{T_{2y}}\}$ Hilbert subspace. We assume  that 
a uniaxial tensile stress has separated the other levels and isolated the ground state triplet manifold. 
Then, as described in appendix~\ref{narrowing}, 
the spectrum narrowing Hamiltonian 
will have the form
\begin{align}
\hat{H}_{sn} =&
\left(q_{\theta}+q_{\eta}\right)
\left(\ket{E_\epsilon}\bra{E_\epsilon} + \ket{T_{2x}}\bra{T_{2x}} + \ket{T_{2y}}\bra{T_{2y}}\right) \nonumber \\
&{}+ \frac{q_\epsilon}{\sqrt{3}}\left(\ket{T_{2x}}\bra{T_{2x}} - \ket{T_{2y}}\bra{T_{2y}}\right).
\end{align}
The form of each $q$ will be
\begin{subequations}
\begin{align}
q_\theta &= -\frac{1}{3}F_0^2\nu_0\left(f_{\perp}^{(2)} - f_{\parallel}^{(2)}\right)\mathcal{E}^2w_\theta, \\
q_\epsilon &= \frac{1}{2}\left[2F_0^2(\nu_0-\nu_1)\left(f_{\perp}^{(2)} - f_{\parallel}^{(2)}\right)\right]\mathcal{E}^2w_\epsilon, \\
q_{\eta} &= \frac{1}{2}F_0^2\nu_1\left(f^{(2)}_{\perp} + f^{(2)}_{\parallel} + \left(f^{(2)}_\perp - f^{(2)}_\parallel\right)n_z^2\right)\mathcal{E}^2,
\end{align}
\end{subequations}
where the parameters $\nu_k$, $f^{(2)}$ and $F_0$ are defined in Section \ref{sec:Variational}.

Because the ground state triplet manifold is well separated from the rest of the states with 
higher energy their contribution to the triplet level shifts is negligible.  
As previously, we eliminate the terms shifting the triplet levels by an equal amount
and set the energy origin at the "center of gravity" of the triplet manifold.
Then the primary effect of spectrum narrowing will consist in renormalization of the Stark 
susceptibility. At the same time, the analogy between electric field and strain remains intact 
within the ground state (i.e. triplet) manifold. The new effective strain variable 
controlling the splitting of the 
triplet levels 
can be 
written as
\begin{gather}
v_\epsilon = e_\epsilon + \frac{1}{2}\frac{\chi'}{\Xi_u}\mathcal{E}^2w_\epsilon
\end{gather} 
where $\chi^\prime$ is the effective susceptibility given by
\begin{align}
\chi^\prime&= \chi + 2F_0^2\left(\nu_0-\nu_1\right)\left(f_\perp^{(2)}-f_\parallel^{(2)}\right) \nonumber \\
& \approx 1.5\,\mathrm{\mu eV(kV/cm)^{-2}}.
\end{align}
While this susceptibility is slightly reduced, we  demonstrate that it is still large enough to induce dramatic shifts of 
electron $g$-factors as well as overall reshaping of ESR lines.

\section{Effects of Zeeman and spin-orbit interactions}
\label{sec:Zeeman-SO}
The full donor electron Hamiltonian,
\begin{equation}
\hat{H}=\hat{H}_{vo}(\bm{\mathcal{E}},e_{ij})+\hat{H}_Z+\hat{H}_{so},\label{eq:H}
\end{equation}
includes the valley-orbit, strain and Stark effects as well as the 
spin-Zeeman and spin-orbit interactions \cite{Watkins:1970wj} characterized 
by $g$-factor anisotropy and by two spin-orbit constants \cite{Watkins:1970wj}
\begin{gather}
g_\perp=1.9984, \quad g_\parallel=1.9994,\\
\lambda_1=2.6 \,\mathrm{\mu eV},\quad \lambda_2=6.9\,\mathrm{\mu eV}.
\end{gather}
\noindent
We can recast the spin-orbit Hamiltonian using  vector operators similar to those introduced by 
Watkins and Ham~\cite{Watkins:1970wj}: 
\begin{equation}
H_{so}
=\frac{1}{2}(\lambda_1\hat{\bm{L}}_1+\lambda_2\hat{\bm{L}}_2)\cdot\hat{\bm{\sigma}},
\end{equation}
where
\begin{align}
\hat{\bm{L}}_1
=&-\frac{i}{2}\sum_{ij}\sum_{ss^\prime}|si\rangle s [
\bm{n}_i
\times
\bm{n}_j
] s^\prime \langle s^\prime j |,\\
\hat{\bm{L}}_2
=&-\frac{1}{2\sqrt{2}}\sum_{ij}\sum_{ss^\prime}|si\rangle \left( [\bm{n}_i\times\bm{n}_j]\cdot\bm{\tau}\right)(\bm{n}_i s - \bm{n}_j s^\prime) \langle s^\prime j |,
\end{align}
where $\bm{\tau}=(1,1,1)$, $\hat{\bm{\sigma}}=\left(\hat{\sigma}_x,\hat{\sigma}_y,\hat{\sigma}_z\right)$,
and $\hat{\sigma}_i$ are the conventional Pauli matrices.
Similarly, the Zeeman 
Hamiltonian can be expressed as:
\begin{equation}
H_{Z}=
\frac{1}{2}g_\bot\mu_B\left[\hat{\bm{\sigma}}\cdot\bm{B}+\varepsilon\sum_{is}|is\rangle
{\hat\sigma_i}B_i \langle is |\right],
\end{equation}
where $\varepsilon=(g_\parallel-g_\perp)/{g_\perp}=5\times 10^{-4}$.

In ESR studies, pulses of ac magnetic field excite donor electron spins 
coupled via the magnetic dipole interaction to a cavity mode with fixed 
frequency $\omega_0$. By varying the strength of the external static magnetic 
field $\bm{B}$, the cavity mode is excited every time when the resonance 
condition,$E_n-E_m=\hbar \omega_0$, is fulfilled.  Here $E_n-E_m$ is a Zeeman 
splitting for the transition between states with predominantly opposite spin 
orientations. The spin-orbit interaction \cite{Watkins:1970wj} gives rise to 
shifts $\hbar \Delta_{nm}$ of the Zeeman splitting for spin-flip transitions
\begin{equation}
E_n-E_m=g_\perp \mu_B B+\hbar \Delta_{nm}.\label{eq:Delta-nm}
\end{equation}
Each transition corresponds to a specific value of static magnetic field $B_{nm}$ at which the resonance with cavity mode is achieved. It is customary to
formally introduce $g$-factors instead of $B_{nm}$ for each transition
\begin{equation}
g_{nm}=\frac{\hbar\omega_0}{\mu_B B_{nm}},\quad E_n-E_m=\hbar \omega_0,\label{eq:gnm}
\end{equation}
where $\mu_B$ is Bohr magneton.
In general, the Zeeman shift $\Delta_{nm}$ is a function of magnetic field magnitude ($B$) and orientation (${\bm \tau}$) as well as of the value of the effective strain ($v_\epsilon$) at the donor location. Introducing a cyclotron frequency, $ \omega$ = $\frac{1}{\hbar}g_\perp\mu_B B$, the values of $g$-factors $g_{nm}$ at each resonance equal
\begin{equation}
g_{nm}=g_\perp\frac{\omega_0}{\omega_{nm}}, \label{eq:gnm0}
\end{equation}
\noindent
where $\omega=\omega_{nm}$ is a root of algebraic equation (cf. (\ref{eq:Delta-nm}))
\begin{equation}
\omega=\omega_{0}-\Delta_{nm}(\omega, {\bm \tau}, v_\epsilon).\label{eq:al}
\end{equation}
\noindent
It is of interest to study the $g$-factor for each transition as function of magnetic field orientation and effective strain:
\begin{equation}
g_{\textrm{nm}}=g_\perp\frac{1}{1-\frac{1}{\omega_0}\Delta_{nm}(\omega_{nm}, {\bm \tau}, v_\epsilon)}.\label{eq:gnm1}
\end{equation}
\noindent
Assuming that the bare Zeeman energy is much greater than the spin-orbit interaction, $\hbar \omega$$\gg$ $\lambda_1,\lambda_2$, the corresponding shifts in Zeeman splitting $\Delta_{nm}(\omega)$ for each transition can be found from the successive orders of perturbation theory corrections to the energy levels $E_n$. As a result, each shift
$\Delta_{nm}$ can be obtained in a form of an expansion in inverse powers of $\omega$:
\begin{equation}
\Delta_{nm}=d_{0}^{nm}+d_{1}^{nm} \omega^{-1}+d_{2}^{nm}\omega^{-2}+d_{3}^{nm} \omega^{-3}+\ldots
\hskip 0.5 cm .\label{eq:d}
\end{equation}
Because the energy corrections contain the powers of bare Zeeman splittings in the
denominator, the expansion (\ref{eq:d}) does not have positive powers of 
$\omega$. The term with zero power of $\omega$ in (\ref{eq:d}) occurs when 
states with the same spin orientation have very close energies and 
coupled by spin-orbit interaction. The latter then splits the Zeeman 
transition frequencies already in zeroth order in $\lambda_{1,2}$ (this 
happens in our case as will see below). Plugging (\ref{eq:d}) into 
(\ref{eq:al}) and inverting the series expansion, one can find a root of 
(\ref{eq:al}) for a given transition, $\omega$=$\omega_{nm}$, in terms of 
inverse powers of $\omega_0$. Then, expressing magnetic field $B_{nm}$ in 
(\ref{eq:gnm}) via $\omega_{nm}$ gives the $g$-factor in the form
\begin{equation}
g=g_\perp \left(1+\frac{d_0}{\omega_{0}}+\frac{d_{0}^{2}+d_{1}}{\omega_{0}^{2}}+\frac{d_{0}^{3}+3d_{0}d_{1}+d_{2}}{\omega_{0}^{3}}+\ldots\right),\label{eq:g}
\end{equation}
where we omitted the state indexes for simplicity. In general, for a given Li 
donor the $g$-factors for individual transitions depend on both the electric field and the strain 
at the donor location through the effective strain variables $v_\theta$, 
$v_\epsilon$, as well as on the magnetic field orientation.

Typically, $g$-factor shifts are very small in the 1$s(A_1)$ ground state of 
substitutional shallow donors in silicon because the spin-orbit interaction only 
involves excited states $1s(E+T_2)$ and the corresponding corrections contain 
very large energy denominators for the valley-orbit splitting between the 
singlet and the rest of $1s$ levels. In the Li donor, the situation is different 
because the bundle of levels, $1s(E+T_2)$, with little or no valley-orbit 
splitting is now a ground state. It is exactly the zero-$th$ order term 
$\propto d_{0}$ in Eq.~(\ref{eq:d}), resulting from this degeneracy, that gives 
rise to very large $g$-factor shifts. Below we will analyze various types of the
spin-flip transitions in the vicinity of the avoided crossing. At the avoided crossing 
the situation is quite involved because the spin-orbit coupling mixes the
states with different orbital characters.  The role of ($xy$)-plane strain or 
electric field is to lift the level degeneracy and to promote direct spin-flip 
transitions between the states with the same orbital characters.

\section{G-factor control with electric fields}
\label{sec:GfacControl}
Here we consider a uniaxial tensile stress along [001] amended by
an electric field along 
[100]. This combination of stress and electric field leads to effective strains $v_\theta$, $v_\epsilon$ such that 
$|v_\theta| >> |v_\epsilon|$. The small, yet nonzero, effective strain 
$v_\epsilon$ splits the triplet levels according to Eq.~(\ref{eq:Vepsilon}). 
If we include the electron spin in the triplet, we have  a sextet of 
spin-orbital electron states. For uniaxial tensile stress $\sigma_{001}\approx 
30$ MPa and magnetic field 0.343 T (corresponding to an ESR cavity mode of 
frequency $\omega_0 = 9600$ MHz), the energy gap separating a ground state 
spin-orbit sextet from the higher-lying states is of the order of 1 meV (cf. 
Fig.~\ref{fig:uniaxial}), and it is much greater than the Zeeman splittings within 
the sextet ($\sim$ 40 $\mu$eV). Therefore, for sub-Kelvin temperatures, the 
system can be described by a reduced Hamiltonian operating on the sextet of 
electron spin states. In what follows, we shall study the effects of 
spin-orbit interaction, small $v_\epsilon$, and magnetic fields upon this 
sextet.

\begin{figure}[t]
{
	\includegraphics[width=3.0in]{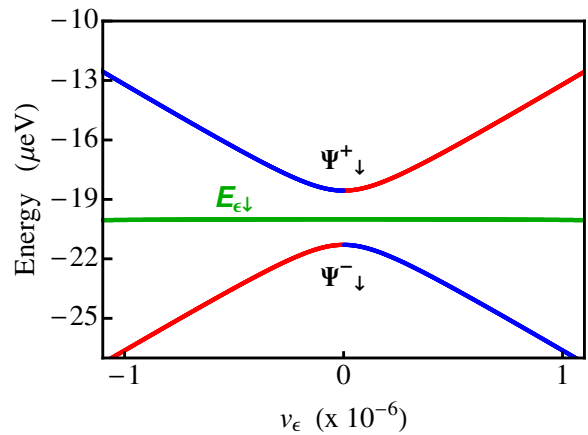}
	\label{fig:NearAvoidedCrossing}
}
\caption{(Color online.) Plot shows the level diagram in the vicinity of one of the avoided 
crossing at $v_\epsilon$=0. Blue, green and red line colors correspond to the 
dominant orbital characters $T_{2y},\,E_\epsilon$, and $T_{2x}$, 
respectively. Symbols $\Psi_{\downarrow}^{\pm}$ indicate the eigenstates 
(\ref{eq:Psiac}) for the corresponding energy levels.
}
\label{fig:triplet-tension}
\end{figure}
\begin{figure}[t]
{
	\includegraphics[width=3.0in]{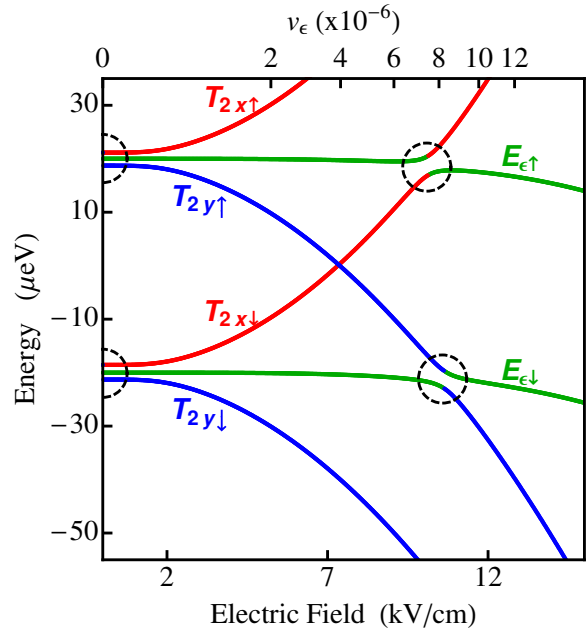}
}
\caption{
(Color online.) Plot shows dependence of the six lowest energy levels on [100] electric field 
with uniaxial tension $\sigma_{001}=30$ MPa and magnetic field $B_{001}=$
0.343 T. 
}
\label{fig:EfieldUniaxial}
\end{figure}

The dependence of the lower triplet of energy levels on the effective strain 
variable $v_\epsilon$ is shown in Fig.~\ref{fig:triplet-tension} for the case 
of a magnetic field in [001] direction. The level structure is symmetric with 
respect to $\pm v_\epsilon$. The behavior of these levels can be understood 
from the fact that when stress and magnetic field are aligned with the same 
crystal axis ($z$) the eigenstates of the donor electron Hamiltonian split into 
two subspaces. The first subspace is formed by the states $\ket{
E_\epsilon, \frac{1}{2}}$, $\ket{T_{2x}, -\frac{1}{2}}$, $\ket{T_{2y}, -
\frac{1}{2}}$ while 
the other one
corresponds to the opposite spin projections 
on magnetic field direction. The states $|\Phi_{\pm}\rangle$ belonging
to different subspaces are  eigenstates 
of the 
operator $\hat Z = \hat R_z(\pi)\hat\sigma_z$, which commutes
with the Hamiltonian. Here
$\hat R_z(\pi)$ is the operator of rotation through angle $\pi$ about the
$z$-axis
and
$\hat Z|\Phi_{\pm}\rangle = \pm|\Phi_{\pm}\rangle$. 
Due to this symmetry, the eigenvalues of $\hat Z$ are good quantum numbers and 
the spin-orbit interaction couples the states within 
each subspace, but not between the subspaces.

For the eigenstates of the donor Hamiltonian, neither the orbital characters ($T_{2x},T_{2y}, E_{\epsilon}$)
nor the spin projections ($\uparrow,\downarrow$) are good quantum numbers. Nonetheless, 
we will
label these eigenstates as $T_{2x\uparrow}$, $E_{\epsilon\downarrow}$ etc., keeping in mind that this is 
{\em predominant}, albeit approximate, character of a given eigenstate.

As shown in Fig.~\ref{fig:EfieldUniaxial}, when a [100] electric field 
increases from zero, the levels $T_{2x}$ and $T_{2y}$ are shifted quadratically 
with the field (linearly with $v_\epsilon$) in opposite directions while the 
level $E_\epsilon$ does not change in the first order in $v_\epsilon$ (its 
variation occurs in the second order from coupling to higher lying states 
$E_\theta,\, A_1$ as $\Delta E = -\eta_\epsilon v_\epsilon^2$ with  
$\eta_\epsilon = 45.\,\mathrm{keV}$). For a given magnetic field and 
effective strain $v_\epsilon=\bar v_\epsilon$ such that
\begin{equation}
\frac{\Xi_u}{\sqrt{3}}\bar v_\epsilon - \eta_\epsilon \bar v_\epsilon^2 
= \hbar\omega,
\label{eq:vbar}\end{equation}
($\bar v_\epsilon\approx 6.3\times10^{-6}$ for $B_{001}=0.343\,\mathrm{T}$), 
there exists an avoided crossing between the levels corresponding to the 
states with dominant characters $\ket{T_{2y}, \frac{1}{2}}$ and $\ket{
E_\epsilon,-\frac{1}{2}}$. Higher in energy by approximately $g\mu_B B$ there 
exists another avoided crossing between $\ket{E_\epsilon,\frac{1}{2}}$ and 
$\ket{T_{2x}, -\frac{1}{2}}$ ($\mathcal{E}\simeq$~11 kV/cm in Fig.~\ref{fig:EfieldUniaxial}). 
For $v_\epsilon\approx-\bar{v}_\epsilon(\omega)$, 
there exists a similar pair of avoided crossings where the orbital character 
$T_{2x}$ is replaced by $T_{2y}$ and vice versa. We note that for the twice 
smaller value of $v_\epsilon\approx3\times10^{-6}$~($\mathcal{E}\simeq$~8 kV/cm), 
the energy levels of the 
states $\ket{T_{2x},-\frac{1}{2}}$, $\ket{T_{2y},\frac{1}{2}}$ undergo a real 
crossing because these states are not coupled to each other by the spin-orbit
interaction.

A different type of avoided-crossing exists for $v_\epsilon\approx 0$. It 
occurs between the levels corresponding to the states with the same spin 
orientation and predominant orbital characters $\ket{T_{2x}}$ and 
$\ket{T_{2y}}$. We note the energies of the states $T_{2x}$, $T_{2y}$, and 
$E_\epsilon$ with the same spin projection are very close to each other. 
Despite this, the state $E_\epsilon$ is not coupled to the other two by the 
spin-orbit interaction. Therefore energy levels of the states with dominant 
characters $\ket{E_\epsilon, \pm \frac{1}{2}}$ are only weakly perturbed by 
spin-orbit interaction and effective strain $v_\epsilon$ (see above).

The level splitting for the the avoided crossing between states with 
opposite spin orientations, $\lambda_2$, is about twice larger than that 
between the states with the same spin orientation (corresponding to 
$\lambda_1$). However, in either case the, level repulsion at the avoided crossings 
is much smaller than the Zeeman energy, and therefore, the analysis of the avoided 
crossings can be done to a leading order within the two-state approximation.
In what follows, we will use the two-state approximation and consider the 
effects of the spin-orbit interactions near the avoided crossings at 
$v_\epsilon = 0$ and $v_\epsilon \approx 6.3 \times 10^{-6}$.

We consider truncated Hamiltonians to describe the avoided-crossings 
at $v_\epsilon$=0. As $E_\epsilon$ is not coupled by the spin-orbit 
interaction to the other states, a two-level Hamiltonian in the basis of the 
corresponding pairs of states $\ket{T_{2x},\frac{1}{2}}$, $\ket{T_{2y},
\frac{1}{2}}$ and $\ket{T_{2x},-\frac{1}{2}}$, $\ket{T_{2y}, -\frac{1}{2}}$ 
is appropriate. The form of these Hamiltonians can be written as
\begin{equation}
\Delta H_{\pm 1/2}^{\rm ac}=\pm \frac{\hbar\omega}{2}\left(\begin{array}{cc}
1 & 0 \\
0 & 1 \\
\end{array}\right)- \frac{\lambda_1}{2} \left(
\begin{array}{cc}
w     & \pm i  \\
\mp i &  -w \\
\end{array}\right),
\end{equation}
where the parameter
\begin{equation}
w= -\frac{2 \Xi_u}{\lambda_1 \sqrt{3}}v_\epsilon \label{eq:w2}
\end{equation}
\noindent
controls the detuning from the avoided-crossing resonance due to  strain 
and/or Stark effects. 
The energies of the pairs of states near the two avoided crossing points equal
\begin{equation}
E_{\pm\frac{1}{2}}^{\sigma}
= \pm \frac{\hbar \omega}{2}+\frac{\sigma\lambda_1}{2}\sqrt{1+w^2}, \label{eq:Eac}
\end{equation}
where $\sigma$=$\pm 1$ is a new {\em orbital} quantum number (in addition to the spin 
$z$-projection $\pm \frac{1}{2}$). The corresponding eigenfunctions are
\begin{equation}
\ket{\Psi^{\sigma},{\pm \frac{1}{2}}} 
=\left(\pm i c_\sigma |T_{2x}\rangle+\sqrt{1-c_{\sigma}^{2}}\,\,|T_{2y}\rangle\right)\otimes\ket{\pm\frac{1}{2}} ,
\label{eq:Psiac}
\end{equation}
where
\begin{equation}
c_\sigma
=-\frac{\sigma}{\sqrt{2}}\left [1-\sigma \frac{w}{\sqrt{1+w^2}}\right ]^{1/2},\qquad \sigma=\pm 1.\label{eq:c}
\end{equation}

It is of interest to consider the magnetic-dipole transitions between states with 
opposite spin orientations near the avoided crossing. The matrix elements of the 
Pauli-matrix $\hat\sigma_x$ for spin-flip transitions between states with 
the same orbital number $\sigma$ equal
\begin{equation}
\bra{\Psi^{\sigma},{-\frac{1}{2}}}\hat\sigma_x\ket{\Psi^{\sigma},{\frac{1}{2}}}
= \frac{\sigma\,w}{\sqrt{1+w^{2}}},\quad \sigma=\pm 1. \label{eq:matel-dir}
\end{equation}
It is seen from (\ref{eq:Eac}) that the frequencies of these transitions (as 
well as that of the transition between the states $|E_\epsilon,\pm 
1/2\rangle$) are very close to $\omega$ forming a triplet of centerlines of 
magnetic dipole transitions. The matrix elements for spin-flip transitions 
between states with opposite orbital numbers $\sigma = \pm 1$ equal
\begin{equation}
\bra{\Psi^{\sigma},{-\frac{1}{2}}}\hat\sigma_x\ket{\Psi^{-\sigma},{\frac{1}{2}}}
= \frac{1}{\sqrt{1+w^{2}}}\label{eq:matel-id}
\end{equation}
According to (\ref{eq:Eac}), the frequencies of these transitions are offset 
from $\omega$ by $\pm \hbar^{-1}\lambda_1 \sqrt{1+w^2}$, forming a doublet of 
satellite lines.

\begin{figure}[t]
{
\includegraphics[width=3.0in]{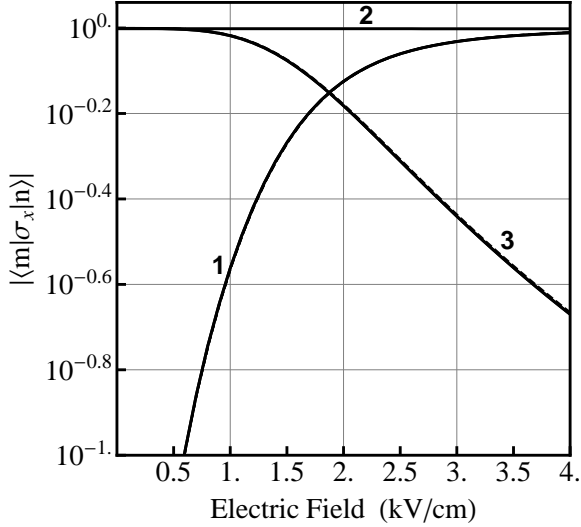}
}
\caption{
Matrix elements for the spin-flip transitions vs. electric field near the 
avoided crossing at $v_\epsilon$=0. Dashed lines (almost undistinguishable in the plots) 
show the analytical results (\ref{eq:matel-dir}),(\ref{eq:matel-id}). The lines marked by:
(1) correspond to matrix elements for the two direct transitions; (2) correspond 
to the matrix element for the transition within the Zeeman doublet 
$|E_\epsilon,\pm 1/2\rangle$ which is very close to 1;  and (3) correspond to the 
matrix elements for the two satellite lines $\Psi_{\uparrow}^{\pm}-
\Psi^{\mp}_{\downarrow}$.}
\label{fig:amplitudePlotsExact}
\end{figure}

Away from the avoided crossing, $|w|\gg1 $, the matrix elements for transitions between states 
of the same $\sigma$ dominate. This behavior 
 can be understood from the fact that, in those regions, each of the 
hybridized orbitals (\ref{eq:Psiac}) is dominated by a single orbital 
character (cf. Fig.~\ref{fig:triplet-tension}(b)). Specifically, when $-w\gg 
1$,
\begin{subequations}
\begin{equation}
\ket{\Psi^{-},{\pm\frac{1}{2}}} \rightarrow \ket{T_{2y},\pm\frac{1}{2}}, \,
\ket{\Psi^{+},{\pm\frac{1}{2}}} \rightarrow \ket{T_{2x},\pm\frac{1}{2}},
\label{eq:cross1}
\end{equation}
and when $w\gg 1$,
\begin{equation}
\ket{\Psi^{-},{\pm\frac{1}{2}}} \rightarrow \ket{T_{2x},\pm\frac{1}{2}}, \,
\ket{\Psi^{+},{\pm\frac{1}{2}}} \rightarrow \ket{T_{2y},\pm\frac{1}{2}}.
\label{eq:cross2}
 \end{equation}
\end{subequations}
Therefore, away from the avoided crossing (i.e. for $|w|\gg 1$) the spin-flip transitions 
(\ref{eq:matel-dir}) conserving $\sigma$ are between states with the same 
dominant orbital character, and the corresponding matrix elements approach 
unity. At the same time, the transitions in (\ref{eq:matel-id}) corresponding 
to satellite lines, are between states with increasingly orthogonal orbital 
characters as $w$ increases and therefore the corresponding matrix elements 
approach zero in this limit.

The behavior of the matrix elements for spin-flip transitions at the 
avoided crossing ($w=0$) is very different than that away from it. It follows 
from (\ref{eq:Psiac}) that at $w$=0 the pairs of states $\ket{\Psi^{+},
-\frac{1}{2}}$,$\ket{\Psi^{-},+\frac{1}{2}}$ and $\ket{\Psi^{-},
-\frac{1}{2}}$, $\ket{\Psi^{+},+\frac{1}{2}}$ are, respectively, symmetric and 
antisymmetric superpositions of the orbital states $i|T_{2x}\rangle$ and 
$|T_{2y}\rangle$:
\begin{subequations}
\begin{eqnarray}
\ket{\Psi^{\mp},{\pm\frac{1}{2}}} 
&=& \frac{i|T_{2x}\rangle+|T_{2y}\rangle}{\sqrt{2}}\otimes\ket{\pm\frac{1}{2}},\label{eq:sym}\\
\ket{\Psi^{\pm},{\pm\frac{1}{2}}} 
&=& \frac{-i|T_{2x}\rangle+|T_{2y}\rangle}{\sqrt{2}}\otimes\ket{\pm\frac{1}{2}}.\label{eq:assym}
\end{eqnarray}
\end{subequations}
The spin-flip transitions (\ref{eq:matel-dir}) that are dominant away form 
avoided crossing $|w|\gg 1$, connecting the same orbital states, become 
suppressed at $w=0$ because they connect  the symmetric and antisymmetric 
superpositions of $i|T_{2x}\rangle$ and $|T_{2y}\rangle$. At the same time, the 
spin-flip transitions (\ref{eq:matel-id}) connecting the states with different 
orbital characters, which are suppressed away from the avoided crossing ($|w|\gg 
1$), become dominant at $w=0$ where they connect the superpositions of 
$i|T_{2x}\rangle$ and $|T_{2y}\rangle$ with the same symmetry. This behavior 
is evident from Fig.~(\ref{fig:amplitudePlotsExact}) giving the 
dependence of matrix elements on electric field (or effective strain 
$v_\epsilon$) obtained using the exact numerical solution for the eigenstates 
of the donor electron Hamiltonian. 

\begin{figure}[htb]
{
\includegraphics[width=2.5in]{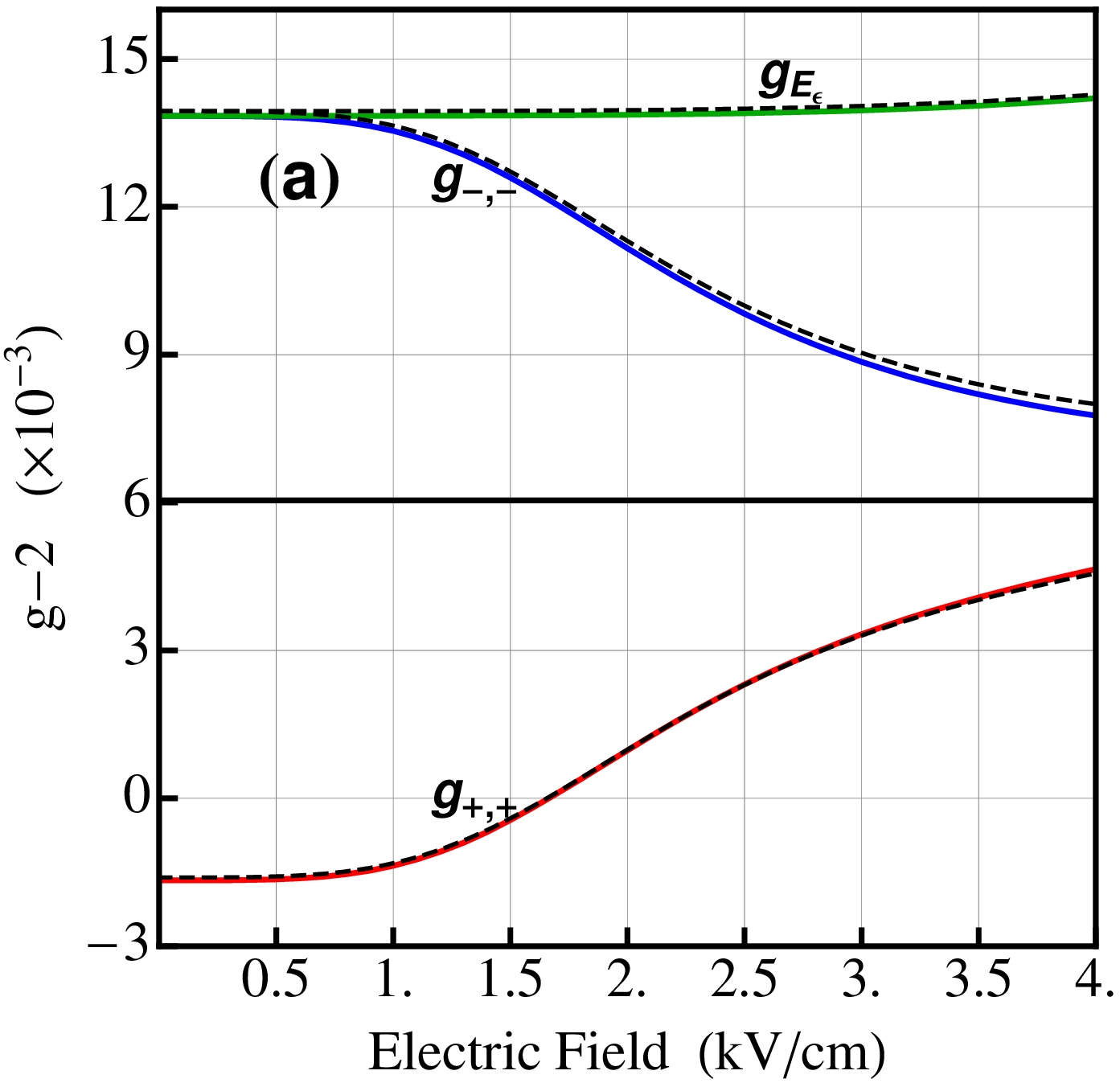}
}
{
\includegraphics[width=2.5in]{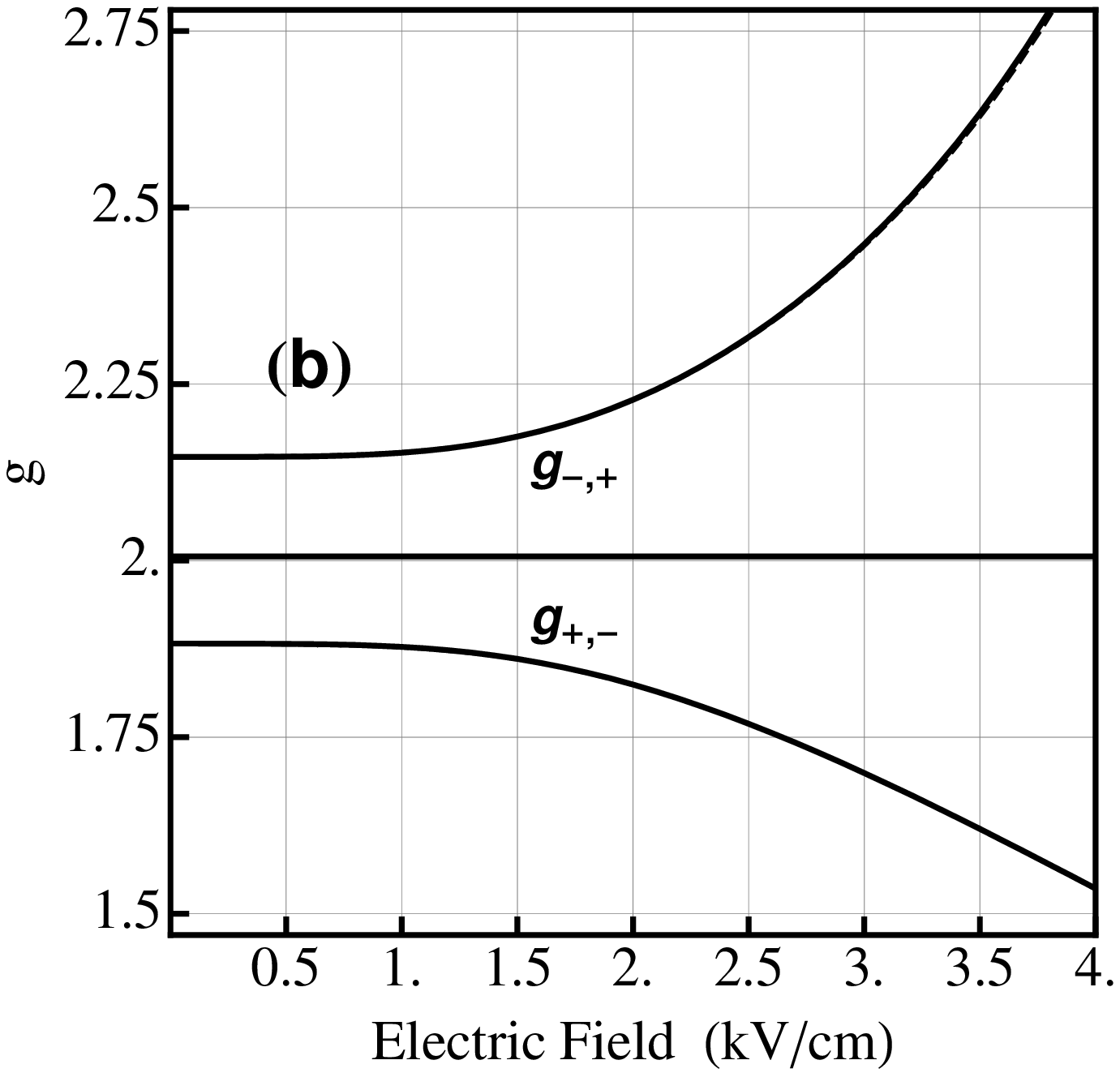}
}
\caption{(Color online.)
$g$-factors vs electric field near the avoided crossing at 
$v_\epsilon$=0. Figure (a) shows three weakly split center lines, 
$g_{++},g_{--}$ and $g_{E_\epsilon}$. Centerline $g_0(1+\delta)$ (\ref{eq:gform}
) is shown in bold. Figure (b) shows the satellite lines $g_{\pm,\mp}$. 
Electric field 3.5 kV/cm corresponds approximately to the strain 
$v_\epsilon\approx10^{-6}$. Dashed lines (unresolved in some plots) 
correspond to analytical approximations given in Eqs.~(\ref{eq:gform}),
(\ref{eq:sat}), (\ref{eq:gppmm}) and (\ref{eq:gE}).
}
\label{fig:tension-g-lines}
\end{figure}

There exist four distinct spin-flip 
transitions between the states of a spin-down doublet $\Psi^{\pm}_{
\downarrow}$ and those of a spin-up doublet $\Psi^{\pm}_{\uparrow }$. We shall 
denote the corresponding $g$-factors as $g_{\sigma, \sigma^\prime}$ ($\sigma,
\sigma^\prime$=$\pm 1$). Based on Eqs.~ (\ref{eq:Eac}) (see also 
Fig.~\ref{fig:triplet-tension}) there are two satellites corresponding to 
$g_{-,+}$, $g_{-,+}$ and two closely spaced center lines corresponding to 
$g_{-,-}$, $g_{+,+}$. There is a third center line with $g$-factor $g_{
E_\epsilon}$ corresponding to the transition between the states $\ket{E_
\epsilon,\pm \frac{1}{2}}$. All $g$-factors can be obtained by numerical 
solution of Eq.~(\ref{eq:al}) for the corresponding transitions. The results 
are shown in Figs.~\ref{fig:tension-g-lines}.

The numerical results can be very closely approximated analytically when 
the ($xy$)-plane strain splitting of $T_{2x},\, T_{2y}$ is much smaller than 
the Zeeman splitting, that is, for $|v_\epsilon|$$\ll$$\bar v_\epsilon$ 
(\ref{eq:vbar}). We find the coefficients in (\ref{eq:d}) by a perturbation 
theory expansion in the spin-orbit interaction constants $\lambda_{1,2}$ using 
the basis of ``correct" states (\ref{eq:Psiac}) in zeroth order. All five 
$g$-factors will have the form 
\begin{subequations}
\begin{gather}
g=g_0(1+\delta)+ \Delta_{\mu}^{\eta},\label{eq:gform} \\
\delta=-\frac{1}{3}\left(\varepsilon\frac{g_\perp}{g_0}\right)+\frac{1}{8}\left(\frac{\lambda_{2}}{\hbar\omega_0}\right)^{2},\label{eq;delta}
\end{gather}
\end{subequations}
where $g_0$ is a singlet $g$-factor and $\Delta_{\psi}^{\eta}$ corresponds to 
different shifts for the various $g$-factors.

For the satellite lines, the shift $\Delta^{\rm sat}_{\sigma}$ equals up to 
the 3$^{\rm rd}$ order in $\lambda_{2}/\omega_0$:
\begin{equation}
\Delta^{\rm sat}_{\sigma}= \sigma\frac{g_0 \lambda_1 D(v_\epsilon)}{\hbar\omega_0-\sigma \lambda_1 D(v_\epsilon)}+\sigma\frac{5}{8}\frac{\lambda_{1}\lambda_{2}^{2}D(v_\epsilon)}{(\hbar\omega_0)^3}.\label{eq:sat}
\end{equation}
Here, we used a secular approximation in (\ref{eq:gnm1}) keeping the terms 
linear in $\lambda_1/\omega_0$ in the denominator because they are proportional to 
the dimensionless parameter $D$, which increases away from the avoided 
crossings
\begin{equation}
D(v_\epsilon)=\sqrt{1+w^2}=\sqrt{1+\frac{4}{3}\left(\frac{\Xi_u}{\lambda_1}\right)^2 v_{\epsilon}^{2}},\label{eq:D}
\end{equation}
where $w$ is given in (\ref{eq:w2}). At the avoided crossing, $g$-factor 
shifts from the centerline $g_0(1+\delta)$ for satellites are near $2\lambda_1/
\hbar\omega_0\approx0.13$ (for $\omega_0= 9600\,\mathrm{MHz}$). As we  
discussed previously, while the $g$-factor shifts for satellite lines 
increase away from the avoided crossing, the strength of the lines decrease. 
Note, however, that with detuning from the avoided crossing corresponding to 
$\mathcal{E}=3.5\,$kV/cm ($v_\epsilon\approx 10^{-6}$) the matrix element is 
only suppressed by a factor of 4 while the $g$-factor shift is already of the 
order of unity. This giant change in $g$-factor can be seen by comparing the 
curves $g_{-,+}$ and $g_{+,-}$ in Fig.~\ref{fig:tension-g-lines}(b), and the curves 
labeled \textrm{3} in Fig.~\ref{fig:amplitudePlotsExact}.

Fig.~\ref{fig:tension-g-lines}(a) shows the $g$-factors $g_{+,+}$, $g_{-,-}$ and 
$g_{E_\epsilon}$ corresponding to three very closely spaced centerlines. Their 
splitting is not resolved within the 2-level picture near the avoided crossing given 
by Eqs.~(\ref{eq:Eac}). We must take into account the transitions 
between the states $|\Psi^{\pm}_{\uparrow,\downarrow}\rangle $ and the 
corresponding states $|E_\epsilon\rangle$ with opposite spin ordination, to determine
the splitting.  The 
values of $\Delta^{\rm c}$ give the shifts of three centerlines from the 
common center. For $g_{\sigma, \sigma}$ ($\sigma$=$\pm$1) they are obtained
from the expansions of $\Delta^{\rm c}_{\sigma}$ in powers of $\omega_{0}^{-1}$:
\begin{equation}
\Delta^{\rm c}_{\sigma}\simeq-\frac{\sigma\lambda_{2}^{2}}{4\hbar^2\omega_{0}^{2}D}+\frac{\lambda_{2}^{2}(D-\sigma)(D^3\lambda_{1}^{2}+(D-\sigma)\lambda_{2}^{2})}{16 D^2\omega_{0}^{4}\hbar^4}.\label{eq:gppmm}
\end{equation}
Here $\sigma=\pm$ and $D$ (\ref{eq:D}) is a dimensionless detuning from 
resonance. Note a slight asymmetry between the line shifts that occurs in the 
$4$th order in $\lambda_2/\hbar\omega_0$. Analytical expressions match very 
closely exact numerical results as demonstrated in Fig.~\ref{fig:tension-g-lines}(a) 
where analytical results are shown with dashed lines. The $g$-factor shift for 
the transitions between the $|E_\epsilon,\pm 1/2\rangle$ states equals
\begin{equation}
 \Delta^{\rm c}_{E_\epsilon}
 =\frac{\lambda_{2}^{2}}{4\hbar^2\omega_{0}^{2}}+\frac{\lambda_{2}^{2}(D^2\lambda_{1}^{2}+2\lambda_{2}^{2})}{8\hbar^4\omega_{0}^{4}}.\label{eq:gE}
\end{equation}
Note that the dependence of this $g$-factor on the effective strain
variable $v_\epsilon$ occurs only in the 4$^{\rm th}$ order in $\lambda_2/
\hbar\omega_0$ leading to its slight decrease with $v_\epsilon$ also shown in 
Fig.~\ref{fig:tension-g-lines}(a). In the leading order, $g$-factor shifts are 
symmetric: $\Delta_{+}^{\rm c}\simeq-\Delta_{-}^{\rm c}$. At the avoided 
crossing ($D=1$), $g$-factor shifts $\Delta^{\rm c}_{-}$ and $\Delta^{\rm c}_
{E_\epsilon}$ nearly coincide as can also be seen in 
Fig.~\ref{fig:tension-g-lines}(a). Away from the avoided crossing $D$$\gg$~1, the lines 
$\Delta_{\pm}^{\rm c}$ asymptotically approach each other as they are shifted down 
by $(\lambda_2/\hbar\omega_0)^2/4\approx0.0076$ from the line $\Delta_{E_
\epsilon}$. We finally note that in the expressions above (\ref{eq:sat}),(
\ref{eq:gppmm}), (\ref{eq:gE}), we have replaced the singlet $g$-factor $g_0$ 
with the integer value 2.

The above predictions can be better understood and visualized if we plot the ESR 
lineshapes for different electric fields. Microwave fields stimulate the spin-flip transitions, 
with resonance emerging as a decrease of the intensity of the received 
microwave field. The ESR lineshape results are predicted through a simple Lorentzian model. 
To account for thermal excitation, we assume
that the total contribution to 
the ESR lineshape from a single donor exposed to microwave light of frequency 
$\omega_0$ will take the form of 
\begin{gather}
\varkappa\left(\bm{B},\bm{\mathcal{E}},\bm{e}\right) = \mathcal{Z}^{-1}\sum_{q,r>q} 
e^{-E_q/k_B T}
\varkappa_{qr}
\left(\bm{B},\bm{\mathcal{E}},\bm{e}\right),
\end{gather}
where $\mathcal{Z} = \sum_n \exp(-E_n/k_B T)$. Here $\varkappa_{qr}$ is the lineshape of a single spin-flip transition between the states 
denoted as $q$ and $r$. The single-transition contribution reads
\begin{gather}
\varkappa_{qr}\left(\bm{B},\bm{\mathcal{E}},\bm{e}\right) = \frac{1}{2\pi}\frac{4\omega_0\Gamma \left|
\chi_{qr}\left(\bm{B},\bm{\mathcal{E}},\bm{e}\right)
\right|^2}{\Gamma^2 + \left[\omega_0 - 
\Delta_{rq}\left(\bm{B},\bm{\mathcal{E}},\bm{e}\right)
\right]^2},
\end{gather}
where $\Delta_{rq} = (E_r - E_q)/\hbar$ and $\chi_{qr} = \bracket{q}{\hat{
\sigma}_x}{r}$ is the spin-flip transition amplitude between these two states. 
The parameter $\Gamma$ expresses the degree of line-broadening due to 
dispersive mechanisms such as thermal effects and spontaneous phonon emission. 

Away from the circled 
avoided  crossings in Fig.~\ref{fig:EfieldUniaxial}, the transitions are strongest between the states of 
the same orbital character. At an electric field of $3\,\mathrm{kV/cm}$, we 
would expect to see three strong absorption lines corresponding to the three 
resonant transitions between $\ket{T_{2x}, \uparrow} \leftrightarrow 
\ket{T_{2x}, \downarrow}$, $\ket{T_{2y}, \uparrow} \leftrightarrow 
\ket{T_{2y}, \downarrow}$, and  $\ket{E_\epsilon, \uparrow} \leftrightarrow 
\ket{E_\epsilon, \downarrow}$.
\begin{figure}[htbp]
{
\includegraphics[width=2.2 in]{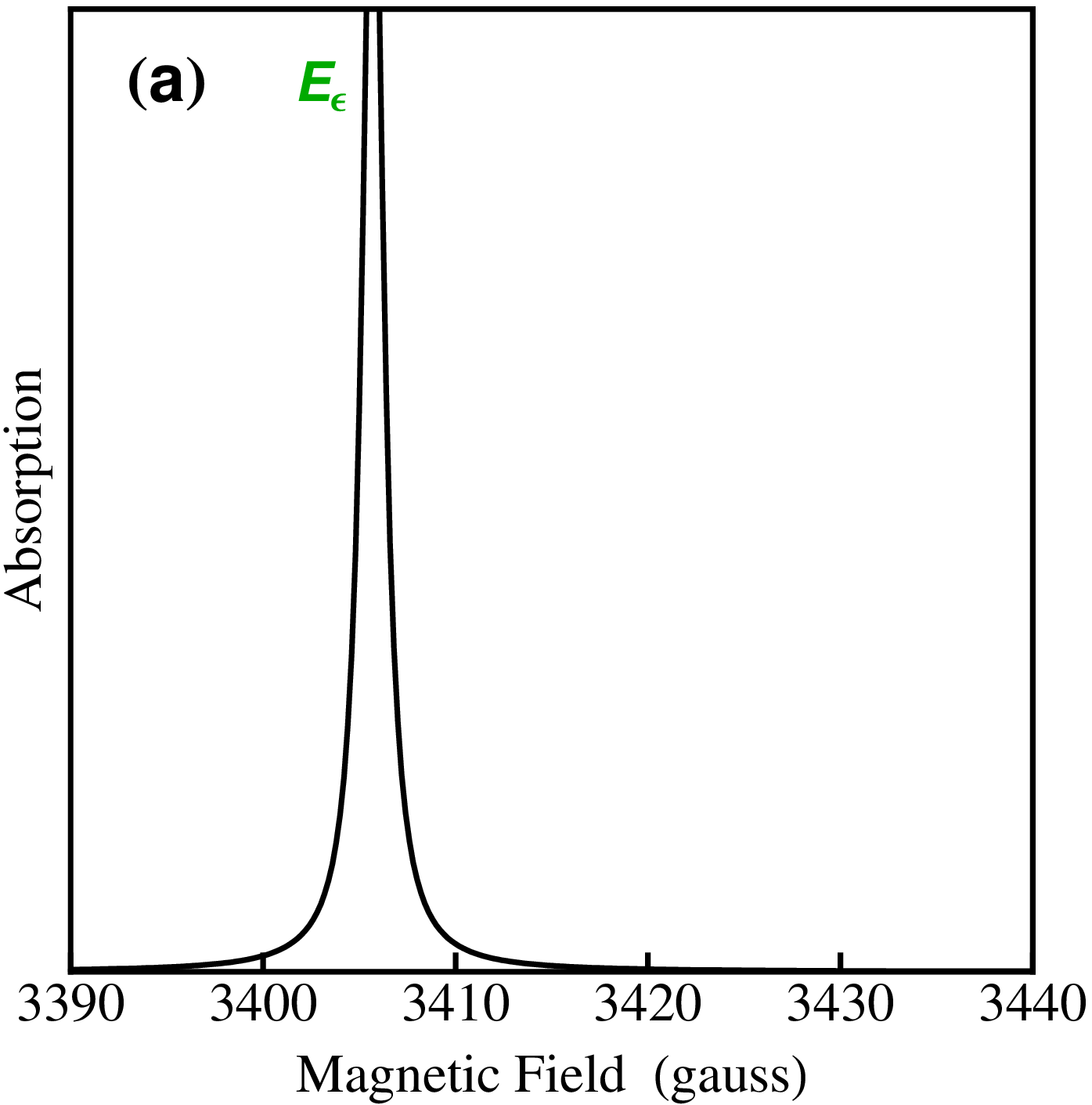}
}
{
\includegraphics[width=2.2 in]{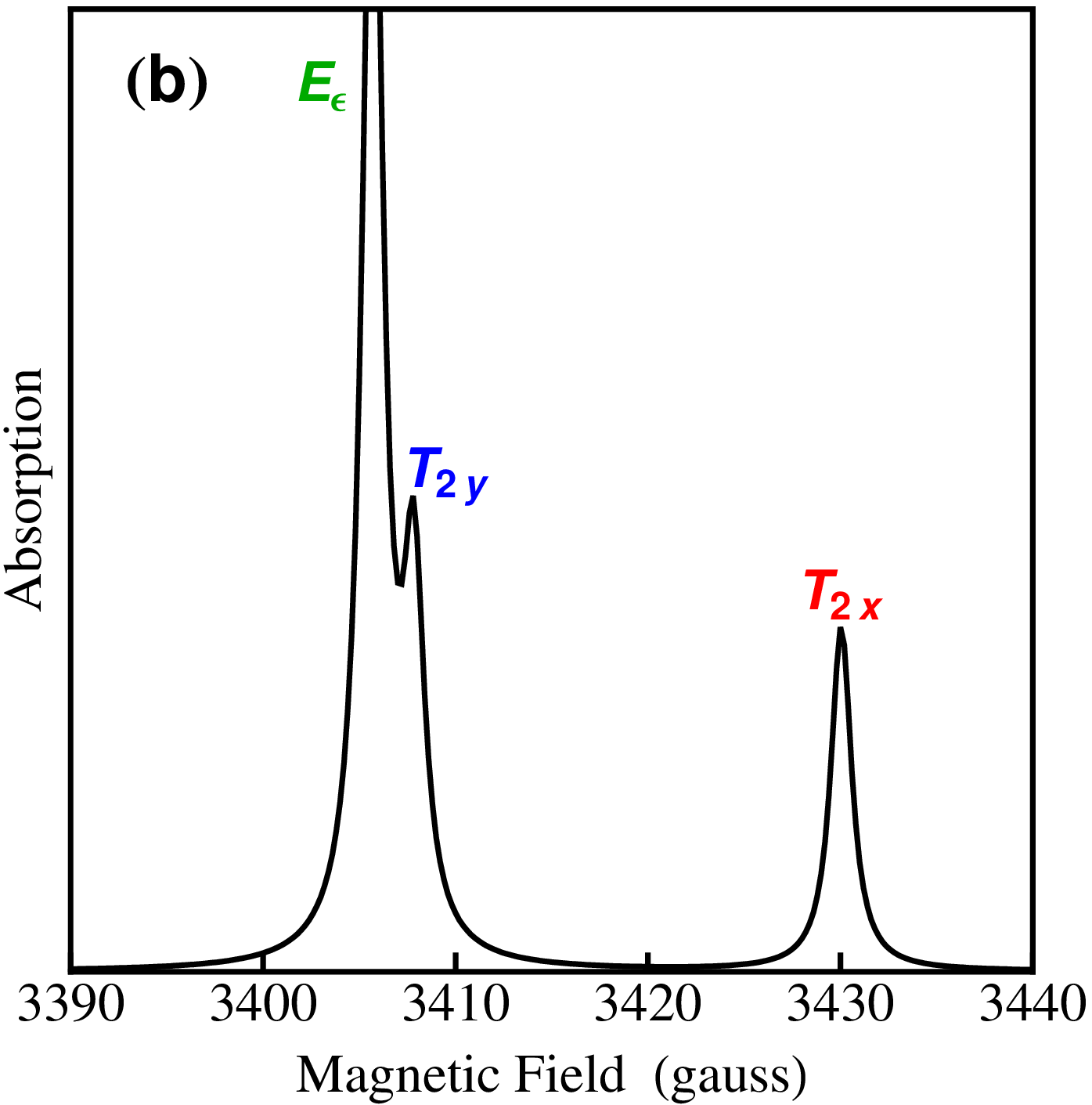}
}
{
\includegraphics[width=2.2 in]{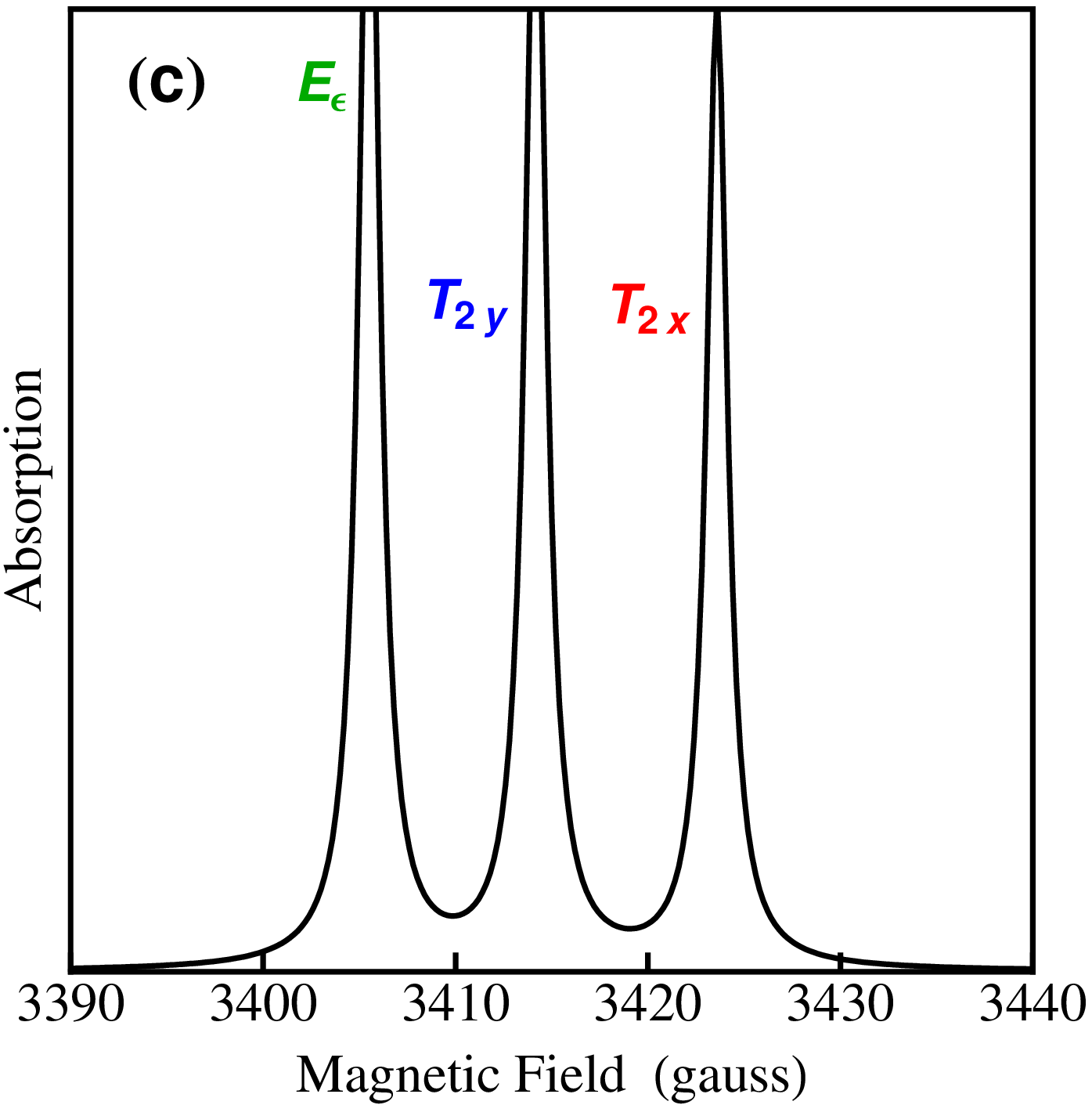}
}
\caption{
(Color online.) Modeled ESR lineshape (9600 MHz cavity mode, 1.3 gauss FWHM, 4 Kelvin) for 
different ${\mathcal E}$-fields along [100]: (a) ${\mathcal E}$=0, (b) ${\mathcal E}$=
1.5~kV/cm, and ${\mathcal E}$=3~kV/cm. The emergence and shifts of  $T_{2x}$ and $T_{2y}$ 
ESR lines are clearly seen. }
\label{fig:lineshapes}
\end{figure}
At the avoided crossing with $v_\epsilon = 0$, a very different situation will 
be in effect. The transition corresponding to $T_{2x}$ and $T_{2y}$ 
is suppressed, and new transitions with greatly shifted $g$-factors will 
emerge. In the above nomenclature, the transitions 
corresponding to the $g$-factors $g_{+,+}$ and $g_{-,-}$ are suppressed 
in favor of satellite transitions corresponding to $g$-factors $g_{+,-}$ and $g_{-,+}$. 
The satellite transitions are very sensitive to the random strains, which implies that
the likelihood of their detection is extremely low.  For this reason we will concentrate on the
centerlines and their electric field dependencies.

The $g$-factor control emerges as the ability to selectively turn 
these lines ``on'' and ``off'', and to shift the lines themselves. 
Fig.~\ref{fig:lineshapes} shows the predictions of our model. The 
suppression  of the two centerlines $T_{2y}$ and $T_{2x}$  at zero field, i.e. near the avoided crossing 
$v_\epsilon=0$,  and their emergence at higher fields can be clearly seen. The zero-field suppression of the 
lines is due to the spin-orbit interaction. The two lines emerge because the electric field shifts the
energy levels away from the avoided crossing where the influence of the spin-orbit coupling is much 
weaker. The overall dependence of the intensity of $T_{2x}$ and $T_{2y}$ lines on the electric field closely
follows that of the spin-flip matrix element shown in Fig.~\ref{fig:amplitudePlotsExact}. As we see in
Fig.~\ref{fig:lineshapes}, in the absence of random strain the Stark effect induces a dramatic shift of the ESR
lines on the order of 10 gauss.

\begin{figure}[tbph]
{
\includegraphics[width=2.5in]{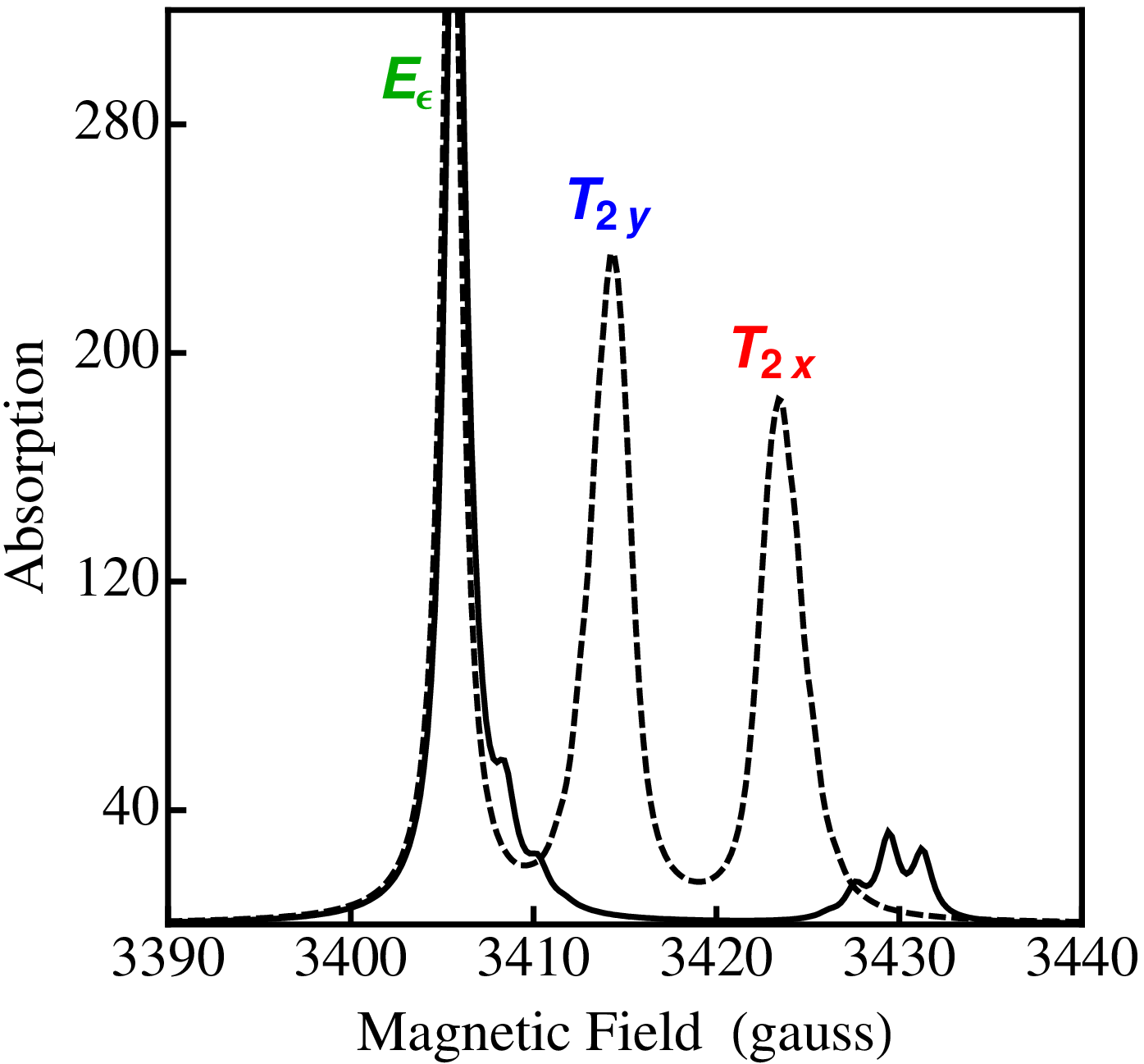}
}
{
\includegraphics[width=2.5in]{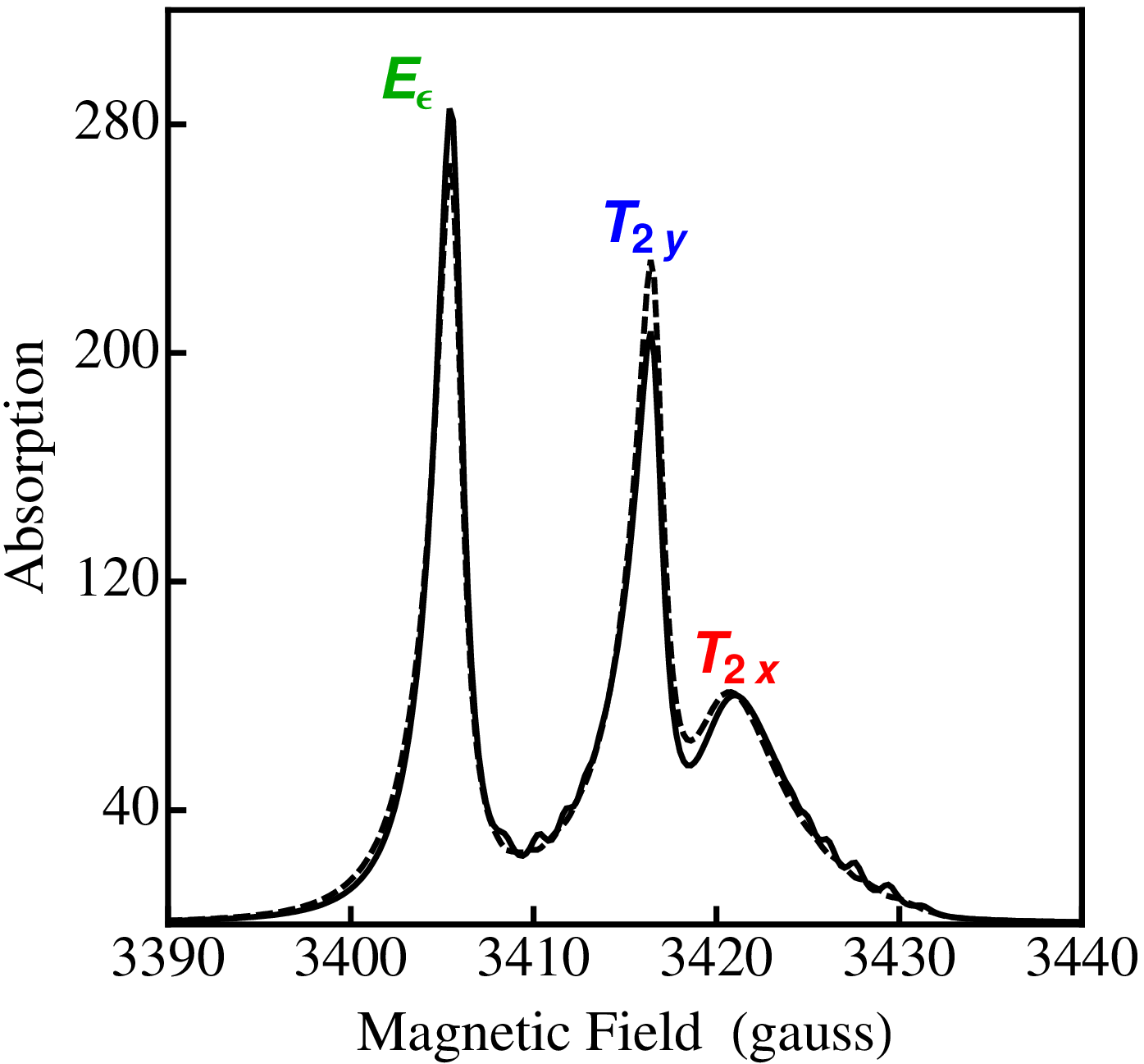}
}
\caption{
(Color online.) ESR absorption lineshapes near $v_\epsilon=0$  for different levels of the random strains and different
electric fields $\mathcal{E}$. The solid and dashed lines represent the 
lineshapes  at $\mathcal{E}=0$ and $\mathcal{E}=$~3 kV/cm respectively. The
Stark shifts are prominent for $\Delta e_\epsilon=10^{-7}$ (top panel) but are almost 
entirely washed out 
for $\Delta e_\epsilon=5\cdot 10^{-7}$ (bottom panel).}
\label{fig:broadenedLines}
\end{figure}

\begin{figure}[tbph]
{
\includegraphics[width=2.5in]{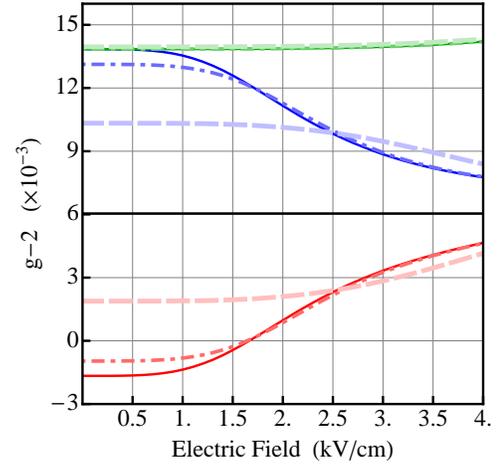}
}
\caption{
(Color online.) Stark shifts of electron $g$-factors for different levels of the random strain. 
The solid lines are the exact $g$-factors at zero
random strain, shown before in Fig.~\ref{fig:tension-g-lines}. The dot-dashed lines 
correspond to $\Delta e_\epsilon=10^{-7}$ and the dashed lines represent the case of 
$\Delta e_\epsilon=5\cdot10^{-7}$.
}
\label{fig:avgG}
\end{figure}

In Figs~\ref{fig:broadenedLines} and \ref{fig:avgG}, we show the effects of random strains upon the ESR 
lineshapes and predicted Stark shifts of Li-donor $g$-factors. We assume that the experimental ESR
signal $\bar{\varkappa}$ can be modeled as an ensemble average over the Gaussian distribution of the
random strains:
\begin{gather}
\bar{\varkappa}\left(\bm{B},\bm{\mathcal{E}}\right) 
= \int n(\bm{e})\varkappa\left(\bm{B},\bm{\mathcal{E}},\bm{e}\right)\,\diff \bm{e}.
\end{gather}
Here $n(\bm{e})$ is the Gaussian distribution function and the integration is taken over 
the strain variables $e_\theta$ and $e_\epsilon$. We further assume that 
a strong uniaxial tensile stress is applied to the sample, and the random internal 
strains shifting $e_\theta$ will have a negligible contribution to the overall ESR signal, i.e. may be safely
ignored. Thus we consider only the effect of random variations of $e_\epsilon$,   assuming
they are described by the Gaussian distributions with different standard deviations
(uncertainties)
$\Delta e_\epsilon$.

To better understand the random strain effects presented in Figs.~\ref{fig:broadenedLines} 
and \ref{fig:avgG}, let us recall our previous finding that the electric field
can strongly affect the Zeeman splittings of the centerlines only in the
vicinity of the avoided crossings (Fig.~\ref{fig:EfieldUniaxial}).
In the domain of random strains with $\Delta e_\epsilon$ not exceeding $10^{-7}$, most
of the donors reside near the avoided crossing $v_\epsilon=0$. Thus, for the majority of these
donors, the centerline transitions $T_{2y}$ and $T_{2x}$ will be suppressed at $\mathcal{E}$=0; however these transitions
will emerge for $\mathcal{E}$-fields exceeding 1 kV/cm.  At $\mathcal{E}$=3 kV/cm, the donor spectra will be shifted from 
$v_\epsilon=0$ to $v_\epsilon$, exceeding the standard deviation $\Delta e_\epsilon$. As a result,
we see pronounced Stark shifts of the ESR lines on the order of 10 gauss (see Fig.~\ref{fig:lineshapes} and
the top panel of Fig.~\ref{fig:broadenedLines}).

The strain disorder randomly shifts the spectra of individual donors
away from the the avoided crossing $v_\epsilon=0$. For broad distributions
most of the donors are subject to the random strains $e_\epsilon$ that exceed $v_\epsilon$
corresponding to $\mathcal{E}$=3 kV/cm. The $g$-factors of these donors are saturated and are insensitive
to the applied $\mathcal{E}$-field (see Fig.~\ref{fig:tension-g-lines} (a)). As a result, the majority of the Li spins do
not display any significant line shifts induced by the electric field, and the Stark features are washed out of the 
average signal of the ensemble.      
As we see from the bottom panel of Fig.~\ref{fig:broadenedLines}, for  $\Delta e_\epsilon=5\cdot 10^{-7}$,
the two lines $T_{2x}$ and $T_{2y}$ 
already exist at $\mathcal{E}$=0, which means  they are induced by the random strain.
The latter also broadens the lines and pins them down at the zero-field positions.  
In the domain of larger random strains 
$\Delta e_\epsilon > 5\cdot 10^{-7}$ the two lines will merge, broaden 
and eventually collapse, resulting in a spectrum  insensitive to the electric field, with one narrow line $E_\epsilon$. 
The electric field dependencies of the average $g$-factors corresponding to different
$\Delta e_\epsilon$ (Fig.~\ref{fig:avgG}) clearly display flattening with increase of $\Delta e_\epsilon$.

The typical values of the random strain uncertainties $\Delta e_\epsilon$, which allow for observation of appreciable Stark shifts
in Li-doped Si, should be in the range of $\Delta e_\epsilon\sim 10^{-7}$ (Fig.~\ref{fig:broadenedLines}). While this requirement is very stringent, the 
possibility of the  growth 
of low-random-strain Si materials
has been demonstrated in photoluminescence experiments~\cite{Yang:2006uk} 
with $^{28}$Si epilayers grown on natural silicon substrates.
Yang {\em et al.}~\cite{Yang:2006uk} were able to resolve splitting of bound exciton lines 
due to the lattice constant mismatch $\Delta a/a\sim 10^{-6}$ between the epilayer and the substrate.  The observed exciton
linewidths are at least an order of magnitude smaller than the splitting, which is indicative of the required level of the random strains $\sim 10^{-7}$. The latter are most likely caused  by isoelectronic impurities  or complexes
(e.g. carbon~\cite{Safonov:1996uu}), which implies that the chemical purity of the material is a key factor in reducing these
strains. 

\section{Summary and Discussion}
\label{sec:Conclusions}
We have developed a theory for the Stark effect for lithium donor spins in silicon.
The anisotropy of the effective mass leads to the anisotropy of the 
quadratic Stark susceptibility, which we determined using the Dalgarno-Lewis 
exact summation method \cite{Dalgarno:1955bi}. The theory is asymptotically 
exact in the field domain below Li-donor ionization threshold, relevant to the 
Stark-tuning ESR experiments \cite{Bradbury:2006gz}. 
Using this theory as a calibration tool we devised a new variational wave function for a shallow
donor in the electric field. The variational function replicates the exact small-field asymptotic results 
and is robust for large fields. 

With the calculated Stark susceptibilities and the new variational function at hand, we  predicted 
and analyzed several important physical effects. First, we observed that the 
energy level shifts due to the quadratic Stark effect are equivalent to 
and can be mapped onto those produced by an external stress \cite{Watkins:1970wj}. 
Second, we demonstrated that the Stark effect anisotropy, combined with unique 
valley-orbit splitting of a Li donor in Si, spin-orbit interaction and 
specially tuned external stress, may lead to a very strong modulation of the 
donor spin $g$-factor by the electric field. Third, we investigated the influence 
of random strains on the $g$-factor shifts and quantified  the random strain 
limits which are necessary to observe the 
ESR-Stark shifts experimentally. 

The ability to control $g$-factors with electric field and/or stress is a crucial component
of many solid-state based quantum computing schemes. 
\cite{Kane:1998wh,Loss:1998ia,Vrijen:2000dk,Jiang:2001de}
Of particular interest for QIP 
proposals is the situation that emerges when 
states with opposite spin-orientations are allowed to mix through the 
spin-orbit interaction leading to the avoided crossings. This has implications not only 
for $g$-factor control but also for long-range inter-donor 
elastic-dipole coupling, suggesting the possibility of controllable 
interactions between isolated donor qubits.~\cite{Smelyanskiy:2005uv} 
Both of these capabilities can be utilized to implement a universal set of gates based on the Ising 
Hamiltonian.~\cite{DeSousa:2004du} This would require ability to grow Si multilayer structures
with a given value of the in-plane strain. The promising technique,
which can be used for practical purposes of  achieving precise control of the uniaxial and/or
biaxial strain, is the growth of Si films on compliant surfaces.~\cite{Yin:2005cv} 
Another interesting possibility is the utilization of piezo-actuators to implement
local stress control of individual impurities.~\cite{Dreher:2011eua}

\begin{acknowledgments}
This work was supported by NASA Cooperative Agreement NNX10AJ58A,
the United States National Security Agency, DOE Grant No. DE-SC0004890,
and the Mines Medal Fellowship (E.~M.~Handberg).
\end{acknowledgments}

\newpage
\begin{appendix}
\section{Derivation of Eqs.~\eqref{eqn:chain}}
\label{Dalgarno}
\begin{widetext}
We seek to solve the differential problem
\begin{equation}
\left(\nabla^2 - 2\pwrt{}{r}\right)\bm{f} - \lambda\left(\spwrt{}{z} - 2\frac{z}{r}\pwrt{}{z}\right)\bm{f} = -\frac{2 m_\perp a_\perp^3}{\hbar^2}\bm{\zeta},
\label{diff_problemA}
\end{equation}
where $\bm{\zeta} = (x, y, za_\parallel/a_\perp)$. First, we use the product rules
\begin{subequations}
\label{eqn:rule1} 
\begin{align}
\nabla^2\left(\cos(\phi)f(r,\theta)\right) &= \cos\phi\left(\nabla^2 - \frac{1}{r^2\sin^2\theta}\right)f(r,\theta), \\ 
\left.\left.\pwrt{}{z}\right(\cos(\phi)f(r,\theta)\right) &=\cos\phi\pwrt{f}{z},
\end{align}
\end{subequations}
and similar rules for the derivatives of $\sin\phi\cdot f(r,\theta)$. If we substitute $\bm{f}$
in the form of Eqs.~\eqref{eqn:fxyz-ansatz} into Eq.~\eqref{diff_problemA} and use
the rules~\eqref{eqn:rule1} we are able to separate  the $\phi$ 
dependence and obtain partial differential equations for 
$f_\parallel(r,\theta)$ and $f_\perp(r,\theta)$ as follows:
\begin{subequations}
\label{eqn:diff_problem}
\begin{align}
\left(\nabla^2 - 2\pwrt{}{r}\right)f_\parallel - \lambda\left(\spwrt{}{z} - 2\frac{z}{r}\pwrt{}{z}\right)f_\parallel
&= -r\cos\theta\label{diff_problemPara}, \\
\left(\nabla^2 - \frac{1}{r^2\sin^2\theta} - 2\pwrt{}{r}\right)f_\perp - \lambda\left(\spwrt{}{z} - 2\frac{z}{r}\pwrt{}{z}\right)f_\perp
&= -r\sin\theta.\label{diff_problemPerp}
\end{align}
\end{subequations}
These equations can be further simplified by employing product rules involving the
associated Legendre polynomials
\begin{equation}
\label{eqn:rule2}
\left(\nabla^2 - \frac{m^2}{r^2\sin^2\theta}\right)f(r)P_l^m(\cos\theta) 
= P_l^m(\cos\theta)\left[\nabla^2 - \frac{l(l+1)}{r^2}\right]f(r),
\end{equation}
and substituting Eqs.~\eqref{eqn:Legendre-expansion} into Eqs.~(\ref{eqn:diff_problem}). This yields:
\begin{subequations}
\label{eqn:sum_l}
\begin{align}
\sum_lP_l(\cos\theta) \left(r^2\hat{D}_r - l(l+1)\right)f_{\parallel,l}
&= \lambda r^2\sum_l\hat{D}_z(f_{\parallel,l}P_l(\cos\theta)) - r^3 \cos\theta
\label{parallel_r} \\
\sum_lP^1_l(\cos\theta)\left(r^2\hat{D}_r - l(l+1)\right) f_{\perp,l}
&= \lambda r^2\sum_l\hat{D}_z(f_{\perp,l}P^1_l(\cos\theta)) + r^3 \sin\theta
\label{perpendicular_r}
\end{align}
\end{subequations}
where $\hat{D}_r$ and $\hat{D}_z$ are defined in Eqs.~\eqref{DrDz}.
\end{widetext}

If $\lambda=0$, Eqs.~(\ref{eqn:sum_l}) have particular solutions 
$f_\parallel=f(r)\cos\theta$ and $f_\bot=f(r)\sin(\theta)$. Indeed, we obtain 
an ordinary differential equation for $f(r)$
\begin{equation}
r^2\frac{d^2f}{dr^2}+2r(1-r)\frac{df}{dr}-2f=-r^3, \label{ode}
\end{equation}
which has the particular solution
\begin{equation}
f(r)=\frac{r(r+2)}{4}. \label{particular}
\end{equation}
The operator $\hat{D}_z$ mixes the Legendre polynomials with different $l$ and does
not allow to separate $r$ and $\theta$. However, the problem can be reduced
to a chain of coupled ordinary differential equations.
To proceed, we employ another set of rules:
\begin{subequations}
\label{eqn:rule3}
\begin{gather}
\pwrt{}{z}f(r)P_l^m(\cos\theta)
= \cos(\theta)P_l^m\pwrt{f}{r} - \frac{f}{r}\sin(\theta)\pwrt{P_l^m}{\theta}, \\
\cos(\theta)P_l^m = \frac{l+m}{2l+1}P_{l-1}^m + \frac{l-m+1}{2l+1}P_{l+1}^m, \\
\sin(\theta)\pwrt{P_l^m}{\theta} = -\frac{(l+1)(l+m)}{2l+1}P_{l-1}^m 
+\frac{l(l-m+1)}{2l+1}P_{l+1}^m.
\end{gather} 
\end{subequations}
We use these rules to rewrite $r^2\hat{D}_zf_l(r)P_l^m(\cos\theta)$ 
as
\begin{align}
r^2\hat{D}_z\left(f_l(r)P_l^m(\cos\theta) \right)
=& \Big(P_{l-2}^m\hat{\alpha}_{l-2}^m + P_l^m\hat{\beta}_l^m \nonumber \\
&\quad+ P_{l+2}^m\hat{\gamma}_{l+2}^m\Big)f_l(r),
\end{align}
where $\hat{\alpha}_l^m,\hat{\beta}_l^m, \hat{\gamma}_l^m$ are second order radial 
differential operators having a common structure of Eq.~\eqref{eqn:alpha-operators}
with the coefficients $\alpha_{l,m}^{(i)}, \beta_{l,m}^{(i)}$ and 
$\gamma_{l,m}^{(i)}$ given explicitly as
\begin{subequations}
\begin{align}
\alpha_{l,m}^{(0)} &= \frac{(l+m+1)(l+m+2)}{(2l+3)(2l+5)} \\
\alpha_{l,m}^{(1)} &= 2l+5 \\
\alpha_{l,m}^{(2)} &= -2(l+3) \\
\alpha_{l,m}^{(3)} &= (l+1)(l+3) 
\end{align}
\end{subequations}
\begin{subequations}
\begin{align}
\beta_{l,m}^{(0)} &= \frac{2l^2+2l-2m^2-1}{(2l-1)(2l+3)} \\
\beta_{l,m}^{(1)} &= 2 \\
\beta_{l,m}^{(2)} &= -1 + \frac{4m^2+1}{2l^2+2l-2m^2-1}\\
\beta_{l,m}^{(3)} &= l(l+1) 
\end{align}
\end{subequations}
\begin{subequations}
\begin{align}
\gamma_{l,m}^{(0)} &= \frac{(l-m)(l-m-1)}{(2l-1)(2l-3)} \\
\gamma_{l,m}^{(1)} &= -2l+3\\
\gamma_{l,m}^{(2)} &= 2(l-2)\\
\gamma_{l,m}^{(3)} &= l(l-2) 
\end{align}
\end{subequations}

We then multiply both sides of Eqs.~\eqref{eqn:sum_l} by $P_l^m(\cos\theta)\sin
\theta$ and integrate $\int_{0}^{\pi} d\theta \sin(\theta)P_l^m(
\cos\theta)(...)$, to eliminate $\theta$-dependence  at the expense of 
mixing $f_l$ with different values of $l$. This procedure leads to
Eqs~\eqref{eqn:chain}. 

\section{
Spectrum Narrowing Effect}\label{narrowing}
The matrix of the central cell potential in the basis of valley-orbitals reads: 
\begin{equation}
H^\prime_{vo} = \begin{pmatrix}
\Delta_{0x} & \Delta_{1x} & \Delta_{2xy} & \Delta_{2xy} & \Delta_{2xz} & \Delta_{2xz} \\
\Delta_{1x} & \Delta_{0x} & \Delta_{2xy} & \Delta_{2xy} & \Delta_{2xz} & \Delta_{2xz} \\
\Delta_{2xy} & \Delta_{2xy} & \Delta_{0y} & \Delta_{1y} & \Delta_{2yz} & \Delta_{2yz} \\
\Delta_{2xy} & \Delta_{2xy} & \Delta_{1y} & \Delta_{0y} & \Delta_{2yz} & \Delta_{2yz} \\
\Delta_{2xz} & \Delta_{2xz} & \Delta_{2yz} & \Delta_{2yz} & \Delta_{0z} & \Delta_{1z} \\
\Delta_{2xz} & \Delta_{2xz} & \Delta_{2yz} & \Delta_{2yz} & \Delta_{1z} & \Delta_{0z}
\end{pmatrix}.
\end{equation}
Transforming this the symmetrized-orbital basis, we find
\begin{equation}
H_{vo} = \begin{pmatrix}
D_A & A           & B             & 0   & 0   & 0   \\
A   & D_{E\theta} & C             & 0   & 0   & 0   \\
B   & C           & D_{E\epsilon} & 0   & 0   & 0   \\
0   & 0           & 0             & D_x & 0   & 0   \\ 
0   & 0           & 0             & 0   & D_y & 0   \\ 
0   & 0           & 0             & 0   & 0   & D_z \\ 
\end{pmatrix}\label{VOorbital}
\end{equation}
Here the energy zero is shifted to the ``center of gravity'' of the manifold.  Explicit expressions for the diagonal
matrix elements are
\begin{subequations}
\begin{align}
D_A=
& \frac{1}{3}\left(\Delta_{1x}+\Delta_{1y}+\Delta_{1z}+4\Delta_{2xy}+4\Delta_{2xz}+4\Delta_{2yz}\right),\label{elementsDiagFirst} \\
D_{E\theta}=
&-\frac{1}{6}(\Delta_{0x}+\Delta_{0y} - 2\Delta_{0z}-\Delta_{1x}-\Delta_{1y}-4\Delta_{1z}\nonumber\\
&-4\Delta_{2xy}+8\Delta_{2xz}+8\Delta_{2yz}), \\
D_{E\epsilon}=
&\frac{1}{6}\left(\Delta_{0x}+\Delta_{0y} - 2\Delta_{0z} + 3\Delta_{1x} + 3\Delta_{1y} -12\Delta_{2xy}\right), \\
D_{x} =
& \frac{1}{3}\left(2\Delta_{0x} - \Delta_{0y} - \Delta_{0z} - 3\Delta_{1x}\right), \\
D_{y} =
& \frac{1}{3}\left(2\Delta_{0y} - \Delta_{0x} - \Delta_{0z} - 3\Delta_{1y}\right), \\
D_{z} = 
&\frac{1}{3}\left(2\Delta_{0z} - \Delta_{0x} - \Delta_{0y} - 3\Delta_{1z}\right).\label{elementsDiagLast}
\end{align}
Similarly, the non-diagonal matrix elements can be expressed as
\begin{align}
A =
& - \frac{\sqrt{2}}{6}\left(\Delta_{0x}+\Delta_{0y}-2\Delta_{0z}+\Delta_{1x}+\Delta_{1y}-2\Delta_{1z}\right. \nonumber\\
&+\left .4\Delta_{2xy}-2\Delta_{2xz}-2\Delta_{yz}\right),\label{elementsOffDiagFirst} \\
B =& \frac{\sqrt{6}}{6}\left(\Delta_{0x}-\Delta_{0y}+\Delta_{1x}-\Delta_{1y}+2\Delta_{2xz} - 2\Delta_{2yz}\right), \\
C =& -\frac{\sqrt{3}}{6}\left(\Delta_{0x}-\Delta_{0y}+\Delta_{1x}-\Delta_{1y}-4\Delta_{2xz} + 4\Delta_{2yz}\right).\label{elementsOffDiagLast}
\end{align}
\end{subequations}
At zero electric field, $F_{x} = F_{y} = F_{x} = F_{0}$. 
Using these values in Eqs. (\ref{sim:delta0})-- (\ref{sim:delta2}), and 
substituting them into the matrix (\ref{VOorbital}),
we find the zero-field valley-orbit Hamiltonian to be 
\begin{gather}
H_{vo}^{(0)} = \begin{pmatrix}
D_{A}^{(0)} & 0 & 0 & 0 & 0 & 0 \\
0 & D_{E}^{(0)} & 0 & 0 & 0 & 0 \\
0 & 0 & D_{E}^{(0)} & 0 & 0 & 0 \\
0 & 0 & 0 & D_{T}^{(0)} & 0 & 0 \\
0 & 0 & 0 & 0 & D_{T}^{(0)} & 0 \\
0 & 0 & 0 & 0 & 0 & D_{T}^{(0)} 
\end{pmatrix},
\end{gather}
where each term on the diagonal is such that
\begin{subequations}
\begin{align}
D_{A}^{(0)} &= F_0^2(\nu_1+4\nu_2), \\
D_{E}^{(0)} &= F_0^2(\nu_1-2\nu_2), \\
D_{T}^{(0)} &= -F_0^2\nu_1.
\end{align}
\end{subequations}
We identify each term on the diagonal with the singlet, doublet, and triplet 
binding energies of the $1s$ donor manifold. Since these values are well known 
from experiment, we can use the constants $\nu_0$, $\nu_1$, $\nu_2$ to 
reproduce the spectrum of any shallow donor in Si. 

The spectrum narrowing Hamiltonian is defined by $H_{sn} = H_{vo} - 
H_{vo}^{(0)}$. To evaluate the extent of spectrum narrowing due to 
displacement of the 
electron away from the donor site, we write
\begin{equation}
F_j = F_0\left(1+\delta_j\right),
\end{equation}
where $\delta_j$ is a function of the field. The values of $\delta_j$ can be 
obtained variationally. Our analysis yields
\begin{gather}
\delta_j = \frac{1}{2}(f_{\perp}^{(2)}-f_{\parallel}^{(2)})\mathcal{E}_j^2 -\frac{1}{2}f_{\perp}^{(2)}\mathcal{E}^2,
\end{gather}
where $\mathcal{E}_j$ is the component of the field lying along the $i$th 
valley. For fields up to $10\,\mathrm{kV/cm}$, the values of $\delta$ remain 
below $0.02$. In the valley-orbit matrix the couplings are $\Delta_{0j} = 
\nu_0 F_0^2(1+2\delta_j)$, $\Delta_{1j} = \nu_1 F_{0}^2(1+2\delta_j)$, 
$\Delta_{2ij} = \nu_2 F_{0}^2(1+\delta_i+\delta_j)$ where the terms to the 
first order in $\delta$ (second order in $\mathcal{E}$) are retained.

To simplify the matrix expressions, we express   for 
our valley-orbit matrix in terms of $\delta_i$. Then the spectrum narrowing Hamiltonian matrix 
can be written as the sum of three simpler 
matrices
\begin{gather}
H_{sn} 
= F_0^2\left(\delta_\theta\hat{U}_{\theta}+\delta_\epsilon\hat{U}_{\epsilon}+\delta_\eta\hat{U}_{\eta}\right)\label{spectrum_narrowing}.
\end{gather}
The parameters $\delta_\mu$, determining the strength of the narrowing, are given by
\begin{subequations}
\begin{align}
\delta_{\theta} 
=& \frac{1}{3}\left(-\delta_x - \delta_y + 2 \delta_z\right),\\
\delta_{\epsilon} 
=& \frac{\sqrt{3}}{3}\left(\delta_x - \delta_y\right),\\
\delta_{\eta}
=&  \delta_x + \delta_y.
\end{align}
\end{subequations}
The component matrices may be expressed in the in the symmetrized-orbital basis 
\begin{subequations}
\begin{align}
\hat{U}_{\theta} =
&(\nu_1 + 4\nu_2)\ket{A_1}\bra{A_1}
+(\nu_0+2\nu_1-4\nu_2)\ket{E_\theta}\bra{E_\theta} \nonumber\\
&{}+(2\nu_0 -3 \nu_1)\ket{T_{2z}}\bra{T_{2z}}+\nu_\alpha\ket{A_1}\bra{E_\theta}+H.C. \nonumber \\
&{}-\nu_0\left(\ket{E_\epsilon}\bra{E_\epsilon}+\ket{T_{2x}}\bra{T_{2x}}+\ket{T_{2y}}\bra{T_{2y}}\right)  \\
\hat{U}_{\epsilon} =
&\sqrt{3}\left(\nu_0-\nu_1\right)\left(\ket{T_{2x}}\bra{T_{2x}}-\ket{T_{2y}}\bra{T_{2y}}\right) \nonumber \\
&{}-v_\theta\ket{E_\theta}\bra{E_\epsilon} +v_\alpha\ket{A_1}\bra{E_\epsilon} + H.C. \\
\hat{U}_{\eta} =
&-\nu_1\left(\ket{T_{2x}}\bra{T_{2x}}+\ket{T_{2y}}\bra{T_{2y}}+\ket{T_{2z}}\bra{T_{2z}}\right) \nonumber\\
&{}+(\nu_1-2\nu_2)\left(\ket{E_\theta}\bra{E_\theta}+\ket{E_\epsilon}\bra{E_\epsilon}\right) \nonumber \\
&{}+(\nu_1+4\nu_2)\ket{A_1}\bra{A_1}
\end{align}
\end{subequations}
with $\nu_\alpha = \sqrt{2}\left(\nu_0+\nu_1+\nu_2\right)$, and $\nu_\theta = 
\nu_0+\nu_1-2\nu_2$.

\end{appendix}

\bibliography{/Users/andre/BibTeX/All}
\bibliographystyle{apsrev}
\end{document}